\documentclass[titlepage,12pt]{article}

\usepackage{graphicx}

\newcommand{\vkr}{\vec{k}_{\parallel}}

\newcommand{\kp}{k_{\perp}}

\newcommand{\ox}{\'o}
\newcommand{\nx}{\'n}

\begin{document}
 %\addtolength{\baselineskip}{0.2\baselineskip}% troche wiekszy odstep miedzy liniami

\begin{center}

{\bf Interface-Localized Mode In Bilayer Film \protect{\linebreak}
Ferromagnetic Resonance Spectrum*}

\vspace{2.0ex}

H.\ Puszkarski** and S.\ Mamica

\vspace{1.0ex}

Surface Physics Division, Faculty of Physics, Adam Mickiewicz
University, \newline 61-614 Pozna{\nx}, ul. Umultowska 85, Poland

\end{center}

\vspace{1.0ex}

PACS numbers: 75.70.-i; 76.50.+g; 75.30.Ds

\vspace{1.0ex}

*To be published in "Horizons in World Physics" Vol. 243 (Nova Science Publishers)

%\vspace{1.0ex}

** Corresponding author: henpusz@main.amu.edu.pl

\vspace{1.0ex}

\begin{center} {\bf ABSTRACT} \end{center}

%\vspace{0.8ex}

%\addtolength{\baselineskip}{0.75\baselineskip}% podwojna interlinia

Ferromagnetic resonance (FMR) in exchange-coupled bilayer films
has been the subject of intensive studies in recent years. From
the experimental viewpoint, a characteristic feature of this FMR
is that  some specimens show single resonance, whereas  others
show double resonance. Moreover, double resonance can exhibit a
regular pattern, in which the high-field (HF) line intensity
surpasses that of the low-field (LF) line, or it can exhibit an
{\it inverted} pattern with the HF line less intense than the LF
line. There is a general agreement that the inverted FMR pattern
occurs when the HF line is an {\it 'optic mode'}, {\it i.e.} an
out-of-phase composition of individual sublayer modes, and the LF
line is an {\it 'acoustic mode'}, or an in-phase mode
composition. The existing theoretical explanations of bilayer
ferromagnetic resonance are, as a rule, based on phenomenological
equation of motion of the magnetization vector. In this paper, we
propose a theory of FMR in ultrathin bilayers based on an
entirely microscopic approach, using the Heisenberg model of
localized spins and assuming that the two ferromagnetic sublayers
are exchange-coupled through their interface. Both the strength
and the sign of this interface coupling ($J_{int}$) is arbitrary
(we admit ferromagnetic or antiferromagnetic interface coupling).
The Hamiltonian contains an exchange energy and a Zeeman energy
terms; the external field is assumed to be applied obliquely to
the film surface. We focus on the effects originating from the
interface coupling, though the system is assumed to exhibit also
{\it pinning} effects, originating from surface anisotropy on the
outer surfaces of the film as well as from intrinsic interface
anisotropy present on internal interfaces. The latter anisotropy
is assumed to consist of two components: uni-directional
($\vec{K}_{int}$) and uni-axial ($D_{int}$). We show that the
resonance spectrum in symmetric bilayer is completely independent
of $J_{int}$, but depends strongly on the applied static field
configuration with respect to the interface normal (angle
$\theta$). A critical angle $\theta_{crit}$ is found to exist (as
in the case of single-layer film) for which the multipeak
spectrum reduces to a single-peak one. This rigorous microscopic
FMR theory does explain the inverted pattern of the bilayer FMR
spectrum by assuming the HF line to correspond to an {\it
in-phase mode, but of interface-localized nature}. The intensity
of such localized mode decreases with growing strength of its
localization at the interface and, when the localization becomes
sufficiently strong, becomes lower than the intensity of the other
mode (which is of the bulk type). This gives a possibility to
explore the HF resonance line corresponding to the
interface-localized mode as a potential source of information
concerning the bilayer interface.

\tableofcontents % spis tresci

%\protect{\pagebreak}

\section{Introduction}\label{rozdz wstep}

\hspace*{\parindent} Magnetic multilayers have recently become
the subject of intensive studies, both theoretical and
experimental, in which special attention is paid to interface
parameters, such as interface exchange coupling or interface
anisotropy. This interest in interface is due to its significant
effect on the properties of the multilayer system as a whole. One
of the key methods of interface investigation is based on
ferromagnetic resonance (FMR). The interface coupling
\cite{J1}--\cite{J3} as well as the interface anisotropy
\cite{D1,D2} in a multilayer can be studied by means of the FMR
spectrum. In anisotropy studies, angle relations play an
important role \cite{Teta1}--\cite{Teta3}. Also, FMR has been
recently used in investigating magnetic particles size
distribution \cite{Biasi02}.

From the experimental viewpoint, the FMR in bilayer films is
characterized by the fact that some specimens show single
resonance, whereas others show double resonance. Moreover, the
double resonance can exhibit a regular pattern, in which the
high-field (HF) line possesses an intensity greater than that of
the low-field (LF) line, or it can exhibit an {\it inverted}
pattern with a HF line less intense than the LF line. According
to a commonly accepted interpretation, the inverted FMR pattern
occurs when the HF line is an out-of-phase composition of the
individual sublayer modes ({\it "optic mode"}) and the LF line is
an in-phase mode composition ({\it "acoustic mode"}). Here,
consensus seems to be due to the circumstance that out-of-phase
modes, naturally, are associated with lower net magnetization;
this is invoked as an explanation of the lower intensity of the
HF resonance line in an inverted pattern. However, a perusal of
the literature shows that this interpretation lacks rigorous
proof; its basic assumption - that between the two lines observed
{\it always} the HF line is out-of-phase and the LF line is
in-phase - has never, to our knowledge, been proved. This
stimulated us to take a closer look at the whole problem.

FMR spectra are commonly interpreted on the basis of macroscopic
theories; in Artman-Layadi theory
\cite{Layadi90}--\cite{Layadi01PRB}, often applied to bilayer
film FMR, sublayer magnetizations are regarded as classical
vectors and appear as such in the key expression for free energy
of interactions between sublayer magnetizations. A microscopic
theory of FMR in thin films has been developed by some authors,
including Ferchmin \cite{Ferchmin62}, Puszkarski \cite{HP67,HP89}
and Wojtczak \cite{Wojtczak69,Castillo98}. The Hamiltonian
considered in their studies includes the exchange interaction
energy, Zeeman energy and the uni-axial anisotropy energy, and is
diagonalized by means of Tyablikov-Bogolyubov method
\cite{Tiablikow}. The resulting eigenvalues correspond to
spin-wave energies, indicating the resonance line positions, and
the intensities of these resonance lines can be found from the
corresponding eigenvectors. Thus, a full theoretical image of FMR
spectrum is obtained by using this method.

In this study, the FMR spectrum in a magnetic bilayer film is
investigated as a function of the interface coupling and the
interface anisotropy, on the basis of the theory developed by
Puszkarski in \cite{HP67,HP89}. The Hamiltonian of
non-homogeneous thin film model is specified in Section
\ref{rozdz model} and diagonalized in Section \ref{rozdz diagon}.
The Hamiltonian of ferromagnetic {\it bilayer} film is
diagonalized in Section \ref{rozdz DW}. Section \ref{rozdz dwa
piki} refers to the bilayer FMR spectra composed of two resonance
lines, often reported in experimental studies; as we mentioned
above, in the most common interpretation the low-intensity line
is related to an {\it optic} mode, and the high-intensity line to
an {\it acoustic} mode. However, the FMR spectra computed in our
model prove this interpretation is not always right. In Section
\ref{rozdz wplyw}, we show that {\it symmetric} bilayer FMR
spectrum does not depend on the interface coupling value, and the
effect of the interface anisotropy on the critical angle
appearance and value is studied in Section \ref{rozdz kryt}.

\section{General planar thin film model}\label{rozdz model}

\subsection{Assumptions}\label{rozdz zalozenia}

% Fig. 1
\begin{figure}
  \centering
  \includegraphics[]{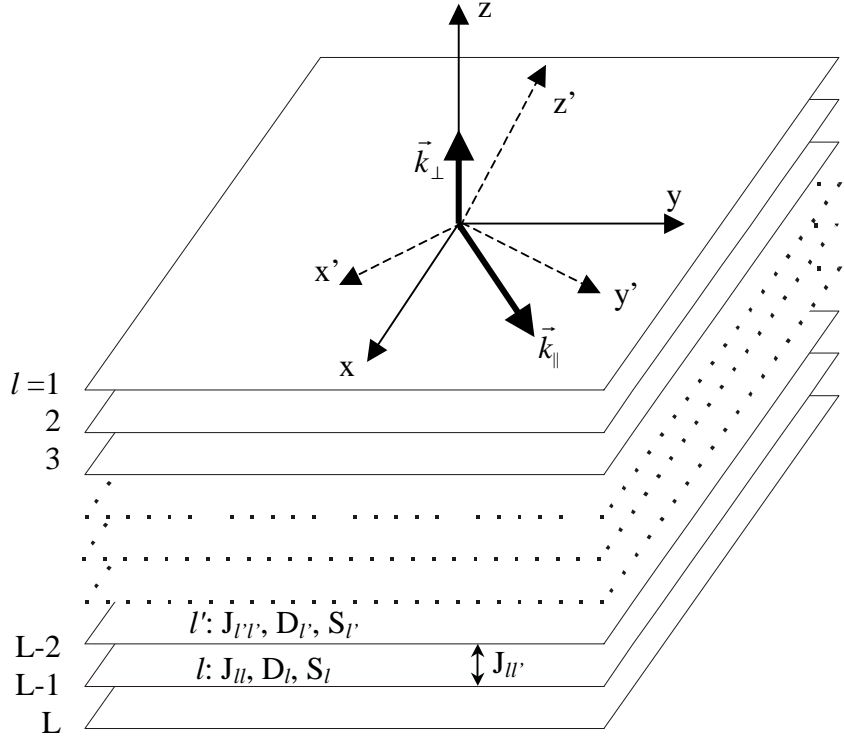}
  %\vspace{11cm}
  \caption[]{The planar thin film model.}\label{rys_model_CW}
\end{figure}

\hspace*{\parindent} In the planar thin film model, referred to
as Valenta model \cite{Valenta57,Valenta62}, the thin film
specimen is assumed to be infinite and homogeneous in the
directions parallel to its surface, but finite and generally
inhomogeneous along the surface normal. In the latter direction,
the inhomogeneous thin film can be divided into a number of
lattice planes parallel to the film surface (see Fig.
\ref{rys_model_CW}). If all the lattice planes are assumed to
have identical crystallographic structure (the same plane lattice
type and equal lattice constant values), all atoms within a plane
have identical neighbourhood, and thus are in identical physical
conditions. A spin is placed in each lattice node. Its position
is defined by the number of the plane to which the spin belongs
($l$) and the site vector within this plane ($\vec{j}$); its
nearest neighbours can lie not only in the same plane or a
neighbouring one, but in farther planes as well (this problem is
more thoroughly discussed in \cite{SM98a,SM98b,SM00a,SM00b}). The
number of plane $l'$ spins being nearest neighbours of a plane
$l$ spin shall be denoted by $z_{ll'}$. Both the spins and the
exchange interactions are assumed to be equal within a lattice
plane (but they can differ from plane to plane).

Two types of coordinate system shall be used below. In the first
one, $xyz$, related to the crystal lattice, $x$ and $y$ axes are
parallel to the film surface, the $z$ axis being oriented along
the surface normal. Another coordinate system to be used,
$x'y'z'$, is related to a single lattice site; in this {\it
local} system, the positive direction of $z'$ axis is defined by
unit vector $\vec{\gamma}_{l\vec{j}}$, indicating the direction
of quantization of the spin in the considered site. The direction
of quantization follows that of the effective magnetic field in
the spin site. Three components contribute to this effective
field: the applied magnetic field, the demagnetization field and
the anisotropy field, the latter usually due to the atomic
magnetic moment interactions with the crystal lattice electric
field (the so-called {\it crystal field}). The assumptions made
within this model imply that the effective field within a lattice
plane is equal in all spin sites ($\vec{H}^{eff}_{l\vec{j}}
\equiv \vec{H}^{eff}_{l}$), and consequently, all the spins
within a single plane have equal direction of quantization
($\vec{\gamma}_{l\vec{j}} \equiv \vec{\gamma}_{l}$).

\subsection{The Hamiltonian}\label{rozdz Ham NCW}

\hspace*{\parindent} We are going to consider a Heisenberg
Hamiltonian of the following form:
\begin{equation}\label{Ham 0}
\widehat{\cal
H}=-2\sum_{l\vec{j};l'\vec{j}'}J_{l\vec{j},l'\vec{j}'}
  \widehat{\vec{S}}_{l\vec{j}}\cdot\widehat{\vec{S}}_{l'\vec{j}'}
  -g\mu_{B}\sum_{l\vec{j}}\vec{H}^{eff}_{l}\cdot\widehat{\vec{S}}_{l\vec{j}}
  -\sum_{l\vec{j}}D_{l}\left(\widehat{S}_{l\vec{j}}^{z}\right)^{2},
\end{equation}
$J_{l\vec{j},l'\vec{j}'}$ denoting the exchange integral, $g$
being the gyromagnetic ratio, and  $\mu_{B}$ being Bohr magneton.
Vector $\vec{j}$ specifies the node position in plane $l$
($\vec{j} \in l$), and vector $\vec{j'}$ plays the same role in
plane $l'$ ($\vec{j'} \in l'$). Symbol
$\sum_{l\vec{j};l'\vec{j}'}$ means the summation involves all the
spin pairs, each pair being considered once only. The three terms
in Hamiltonian (\ref{Ham 0}) correspond to the exchange
interactions energy, the Zeeman energy and the single-ion
anisotropy energy, respectively. The dynamic dipolar fields shall
not be taken explicitly into consideration here, since -- as it
has been shown by Krawczyk \cite{Krawczyk02} -- their effect on
the localized state {\it existence} conditions is minor.

Assuming that the exchange interactions occur between the nearest
neighbours only, and admitting that these interactions, though
equal within a lattice plane, can be different between spins
lying in {\it different} planes (in this case, $J_{l\vec{j},
l'\vec{j}'} = J_{ll'}$), we obtain the following form of the
Hamiltonian:
\begin{equation}\label{Ham}
\widehat{\cal H}= -\sum_{l,\vec{j} \neq l',\vec{j}'}J_{ll'}
  \widehat{\vec{S}}_{l\vec{j}}\cdot\widehat{\vec{S}}_{l',\vec{j}'}
  -g\mu_{B}\sum_{l,\vec{j}}\vec{H}^{eff}_{l}\cdot\widehat{\vec{S}}_{l\vec{j}}
  -\sum_{l,\vec{j}}D_{l}\left(\widehat{S}_{l\vec{j}}^{z}\right)^{2}.
\end{equation}

In order to find the {\it inhomogeneous thin film} eigenstates,
Hamiltonian (\ref{Ham}) must be diagonalized.

\section{Hamiltonian diagonalization procedure}\label{rozdz diagon}

\hspace*{\parindent} Four transformations shall be performed
\cite{HP71}, leading to the diagonal form of Hamiltonian
(\ref{Ham}): 1) transformation to the local coordinate system, 2)
transformation to the boson operators, 3) Fourier transformation
in the film plane and 4) the transformation along the film
surface normal.

\subsection{Transformation to the local coordinate system}

\hspace*{\parindent} First we shall transform Hamiltonian
(\ref{Ham}) from its form in the crystal lattice coordinate
system, $xyz$, to that in the local coordinate system, $x'y'z'$.
This transformation reads \cite{Tiablikow}:
\begin{equation}\label{S prim}
\widehat{\vec{S}}_{l\vec{j}}= \vec{\gamma}_{l\vec{j}}
\widehat{S'}_{l\vec{j}}^{z}+ \frac{1}{\sqrt{2}}
\left(\vec{A}_{l\vec{j}} \widehat{S'}_{l\vec{j}}^{+} +
\vec{A}_{l\vec{j}}^{\ast}\widehat{S'}_{l\vec{j}}^{-}\right),
\end{equation}
$\vec{\gamma}_{l\vec{j}}$ denoting the previously introduced
quantization axis unit vector, and  $\vec{A}_{l\vec{j}}$ having
the following coordinates (provided that $y'$ lies in plane $xy$):
\begin{eqnarray}\label{A wek}
A_{l\vec{j}}^{x}&=&-\frac{1}{\sqrt{2}} \sqrt{1- \left(
\gamma_{l\vec{j}}^{z} \right)^{2}} \left( \gamma_{l\vec{j}}^{x}
\gamma_{l\vec{j}}^{z} + i\gamma_{l\vec{j}}^{y} \right),\nonumber\\
A_{l\vec{j}}^{y}&=&-\frac{1}{\sqrt{2}} \sqrt{1- \left(
\gamma_{l\vec{j}}^{z} \right)^{2}} \left( \gamma_{l\vec{j}}^{y}
\gamma_{l\vec{j}}^{z} - i\gamma_{l\vec{j}}^{x} \right),\\
A_{l\vec{j}}^{z}&=&\frac{1}{\sqrt{2}} \sqrt{1- \left(
\gamma_{l\vec{j}}^{z} \right)^{2}}.\nonumber
\end{eqnarray}
Vectors $\vec{\gamma}_{l\vec{j}}$ and $\vec{A}_{l\vec{j}}$
fulfill the following relations:
\begin{equation}\label{wlasnosci A}
\begin{array}{lll}
\vec{\gamma}_{l\vec{j}}=\vec{\gamma}_{l\vec{j}}^{\ast},
 & \vec{\gamma}_{l\vec{j}}\cdot\vec{\gamma}_{l\vec{j}}=1,
 & \vec{A}_{l\vec{j}}^{\ast}\cdot\vec{A}_{l\vec{j}}=1, \\
\vec{A}_{l\vec{j}}\cdot\vec{\gamma}_{l\vec{j}}=0,
 & \vec{A}_{l\vec{j}}^{\ast}\cdot\vec{\gamma}_{l\vec{j}}=0,
 & \vec{A}_{l\vec{j}}\cdot\vec{A}_{l\vec{j}}=0, \\
\vec{\gamma}_{l\vec{j}}\times\vec{A}_{l\vec{j}}=i\vec{A}_{l\vec{j}},
 & \vec{A}_{l\vec{j}}\times\vec{A}_{l\vec{j}}^{\ast}=
   i\vec{\gamma}_{l\vec{j}}. &
\end{array}
\end{equation}

\subsection{Transformation to second quantization operators}

\hspace*{\parindent} In the second step, the spin operators
(expressed in $x'y'z'$ coordinates) shall be replaced with boson
operators by means of Holstein-Primakoff transformation. In the
resulting Hamiltonian, only quadratic terms shall be retained,
and all the terms of other orders shall be omitted (however, the
commutation rules should be taken into consideration, and
attention paid to the sequence of boson operators when omitting
the higher-order terms).

Holstein-Primakoff transformation reads \cite{Tiablikow}:
\begin{equation}\label{HolPri}
\begin{array}{l}
\begin{array}{lll}
\widehat{S'}_{l\vec{j}}^{+} = \sqrt{2S_{l\vec{j}}}
\widehat{f}_{l\vec{j}} \widehat{a}_{l\vec{j}}, &
\widehat{S'}_{l\vec{j}}^{-} = \sqrt{2S_{l\vec{j}}}
\widehat{a}_{l\vec{j}}^{+} \widehat{f}_{l\vec{j}}, &
\widehat{S'}_{l\vec{j}}^{z} = S_{l\vec{j}}-
\widehat{a}_{l\vec{j}}^{+}\widehat{a}_{l\vec{j}},
\end{array} \\
\,\; \widehat{f}_{l\vec{j}} = \sqrt{1-\widehat{a}_{l\vec{j}}^{+}
\widehat{a}_{l\vec{j}}/2S_{l\vec{j}}} ,
\end{array}
\end{equation}
with operators of creation ($\hat{a}_{l\vec{j}}^{+}$) and
annihilation ($\hat{a}_{l\vec{j}}$) satisfying the following
commutation rules:
\begin{eqnarray}\label{komut a}
[\hat{a}_{l\vec{j}}, \hat{a}_{l'\vec{j'}}^{+}] =
\delta_{ll'}\delta_{\vec{j}\vec{j'}} , &
[\hat{a}_{l\vec{j}},\hat{a}_{l'\vec{j'}}]=0 . &
\end{eqnarray}

Using the approximation of {\it quasi-saturation}, we obtain:
\begin{equation}\label{Trans 1}
\begin{array}{lll}
\widehat{S'}_{l\vec{j}}^{+} =
\sqrt{2S_{l\vec{j}}}\widehat{a}_{l\vec{j}} , &
\widehat{S'}_{l\vec{j}}^{-} =
\sqrt{2S_{l\vec{j}}}\widehat{a}_{l\vec{j}}^{+} , &
\widehat{S'}_{l\vec{j}}^{z} = S_{l\vec{j}}-
\widehat{a}_{l\vec{j}}^{+}\widehat{a}_{l\vec{j}} .
\end{array}
\end{equation}
The above transformations shall be applied to the first two terms
in the Hamiltonian, {\it i.e.} the exchange (bi-ion) interaction
term and the Zeeman (single-ion, linear) term. The
transformations to be applied to the remaining anisotropy term
(single-ion, non-linear), are as follows \cite{HP79}:
\begin{equation}\label{Trans 2}
\begin{array}{l}
\widehat{S'}_{l\vec{j}}^{-}\widehat{S'}_{l\vec{j}}^{+} =
2S_{l\vec{j}}
  \widehat{a}_{l\vec{j}}^{+}\widehat{a}_{l\vec{j}} ,\\
\widehat{S'}_{l\vec{j}}^{+}\widehat{S'}_{l\vec{j}}^{-} =
2S_{l\vec{j}} +
  \left(2S_{l\vec{j}}-2 \right) \widehat{a}_{l\vec{j}}^{+}\widehat{a}_{l\vec{j}} ,\\
\left( \widehat{S'}^{z}_{l\vec{j}} \right)^{2}=S_{l\vec{j}}^{'2} -
  \left(2S_{l\vec{j}}-1 \right) \widehat{a}_{l\vec{j}}^{+}\widehat{a}_{l\vec{j}} ,\\
\widehat{S'}_{l\vec{j}}^{+}\widehat{S'}_{l\vec{j}}^{z} =
  \sqrt{2S_{l\vec{j}}}
  \left(S_{l\vec{j}}-1 \right) \widehat{a}_{l\vec{j}} ,\\
\widehat{S'}_{l\vec{j}}^{z}\widehat{S'}_{l\vec{j}}^{+} =
  \sqrt{2S_{l\vec{j}}}
  S_{l\vec{j}} \widehat{a}_{l\vec{j}} ,\\
\widehat{S'}_{l\vec{j}}^{-}\widehat{S'}_{l\vec{j}}^{z} =
  \sqrt{2S_{l\vec{j}}}
  S_{l\vec{j}} \widehat{a}_{l\vec{j}}^{+} ,\\
\widehat{S'}_{l\vec{j}}^{z}\widehat{S'}_{l\vec{j}}^{-} =
  \sqrt{2S_{l\vec{j}}}
  \left(S_{l\vec{j}}-1 \right) \widehat{a}_{l\vec{j}}^{+} ,\\
\widehat{S'}_{l\vec{j}}^{+}\widehat{S'}_{l\vec{j}}^{+} =
  \sqrt{2S_{l\vec{j}}
  \left(2S_{l\vec{j}}-1 \right)} \widehat{a}_{l\vec{j}}\widehat{a}_{l\vec{j}}  ,\\
\widehat{S'}_{l\vec{j}}^{-}\widehat{S'}_{l\vec{j}}^{-} =
  \sqrt{2S_{l\vec{j}}
  \left(2S_{l\vec{j}}-1 \right)} \widehat{a}_{l\vec{j}}^{+}\widehat{a}_{l\vec{j}}^{+}  .\\
\end{array}
\end{equation}

After transformations (\ref{S prim}), (\ref{Trans 1}) and
(\ref{Trans 2}) Hamiltonian (\ref{Ham}) becomes:
\begin{equation}\label{Ham boz}
\widehat{\cal H}= E_{0}(\vec{\gamma}_{l})+
\sum_{l,\vec{j};l',\vec{j}'}\left(P_{ll'}^{\vec{j}\vec{j'}}
\hat{a}_{l\vec{j}}^{+}\hat{a}_{l'\vec{j'}}
+\frac{1}{2}Q_{ll'}^{\vec{j}\vec{j'}}
\hat{a}_{l\vec{j}}\hat{a}_{l'\vec{j'}}
+\frac{1}{2}Q_{ll'}^{\ast\vec{j}\vec{j'}}
\hat{a}_{l\vec{j}}^{+}\hat{a}_{l'\vec{j'}}^{+}\right)+
\sum_{l,\vec{j}}\left(R_{l}\hat{a}_{l\vec{j}} +
R_{l}^{\ast}\hat{a}_{l\vec{j}}^{+}\right) ,
\end{equation}
where:
\begin{eqnarray}\label{E0}
E_{0}(\vec{\gamma}_{l}) &=& -N\sum_{ll'}z_{ll'}J_{ll'}S_{l}S_{l'}
\vec{\gamma}_{l}\cdot\vec{\gamma}_{l'}-Ng\mu_{B}\sum_{l}S_{l}
\vec{H}^{eff}_{l}\cdot\vec{\gamma}_{l} \nonumber \\
 & & -N\sum_{l}D_{l}S_{l}\left(S_{l}\cos^{2}\theta +
\frac{1}{2}\cos^{2}\theta-\frac{1}{2}\right) ,
\end{eqnarray}
\begin{equation}\label{Plj}
P_{ll'}^{\vec{j}\vec{j'}}= \left\{
\begin{array}{ll}
  2\sum_{n}z_{ln}J_{ln}S_{n}\vec{\gamma}_{l}\cdot\vec{\gamma}_{n}
  +g\mu_{B}\vec{H}^{eff}_{l}\cdot\vec{\gamma}_{l} & \\
  \hspace*{6em} -D_{l}(S_{l}-\frac{1}{2})(1-3\cos^{2}\theta) &
 $ for $ l\vec{j}=l'\vec{j'} , \\
 & \\*[-1.5ex]
  -2\sqrt{S_{l}S_{l'}}J_{ll'}A_{l}^{\ast}\cdot A_{l'} &
 $ for $ l\vec{j}\neq l'\vec{j'} ,
\end{array} \right.
\end{equation}
\begin{equation}\label{Qlj}
Q_{ll'}^{\vec{j}\vec{j'}}=\left\{
\begin{array}{ll}
  -D_{l}\sqrt{S_{l}(S_{l}-\frac{1}{2})}(1-\cos^{2}\theta) &
 $ for $ l\vec{j}=l'\vec{j'} , \\
 & \\*[-1.5ex]
  -2\sqrt{S_{l}S_{l'}}J_{ll'}A_{l}\cdot A_{l'} &
 $ for $ l\vec{j}\neq l'\vec{j'} ,
\end{array} \right.
\end{equation}
\begin{eqnarray}\label{Rl}
R_{l} & = & -2 \sqrt{S_{l}} \sum_{n}z_{ln} J_{ln} S_{n} A_{l}
\cdot \vec{\gamma}_{n} - g \mu_{B} \sqrt{S_{l}} \vec{H}^{eff}_{l}
\cdot A_{l} \nonumber \\
 & & -\frac{1}{2} \sqrt{2S_{l}} (2S_{l}-1) \cos\theta \sqrt{1-\cos^{2}\theta}.
\end{eqnarray}
Symbol $\theta$, appearing in the above equations, denotes the
angle between the film magnetization vector and the surface
normal; $z_{ll'}$, as previously defined, is the number of a
plane $l$ spin nearest neighbours in plane $l'$. Equality
$J_{ll'}=J_{l'l}$ implies that $P^{\vec{j'}\vec{j}}_{l'l} =
P^{\ast\vec{j}\vec{j'}}_{ll'}$ and $Q^{\vec{j'}\vec{j}}_{l'l} =
Q^{\vec{j}\vec{j'}}_{ll'}$, which means that Hamiltonian
(\ref{Ham boz}) is Hermitian.

\subsection{Bi-linear Hamiltonian}

\hspace*{\parindent} As mentioned above, only quadratic terms
shall be retained in the considered Hamiltonian:
\begin{equation}\label{Ham bil}
\widehat{\cal H}=
\sum_{l,\vec{j};l',\vec{j}'}\left(P_{ll'}^{\vec{j}\vec{j'}}
\hat{a}_{l\vec{j}}^{+}\hat{a}_{l'\vec{j'}}
+\frac{1}{2}Q_{ll'}^{\vec{j}\vec{j'}}
\hat{a}_{l\vec{j}}\hat{a}_{l'\vec{j'}}
+\frac{1}{2}Q_{ll'}^{\ast\vec{j}\vec{j'}}
\hat{a}_{l\vec{j}}^{+}\hat{a}_{l'\vec{j'}}^{+}\right).
\end{equation}
The terms of order zero, shifting uniformly the energy scale, are
omitted. The first-order terms vanish when the direction of
$\vec{\gamma}$ corresponds to the minimum energy \cite{HP79}. The
fourth-order terms are related to the interactions between
magnons, and their omitting is justified by our previous
assumptions.

The general method of quadratic form diagonalization was proposed
by Tyablikov and Bogolyubov, and applied to thin film and bilayer
film by Puszkarski \cite{HP68,HP79}. The procedure, analogical to
that used by Ferchmin \cite{Ferchmin62} and Corciovei
\cite{Corciovei63}, is based on two Fourier transformations,
performed in the film plane and along the direction normal to the
film surface. By these operations, boson operators
$\hat{a}_{l\vec{j}}^{+}$ and $\hat{a}_{l'\vec{j'}}$, originally
expressed in the direct space, $l\vec{j}$, are transformed into
the reciprocal space, $\vkr\kp$, in which the Hamiltonian is
diagonal.

\subsection{Fourier transformation in the film plane}

\hspace*{\parindent} The transformation in the film plane reads
\cite{Wojtczak69,Ferchmin62,Corciovei63}:
\begin{eqnarray}\label{op a+a}
\hat{a}_{l\vec{j}}=\frac{1}{\sqrt{N}}\sum_{\vkr } \exp(-i\vkr
\cdot\vec{j})\hat{b}_{\vkr l} , && \\
\hat{a}_{l\vec{j}}^{+}=\frac{1}{\sqrt{N}}\sum_{\vkr } \exp(i\vkr
\cdot\vec{j})\hat{b}_{\vkr l}^{+} , && \nonumber
\end{eqnarray}
$N$ denoting the total number of spins in a single lattice plane
parallel to the surface, and $\vkr =[k_{x},k_{y}]$ being a vector
from the two-dimensional Brillouin zone (coordinates $k_{x}$ and
$k_{y}$ are quantized through imposing Born-K\'arm\'an cyclic
boundary conditions in the $x$ and $y$ directions). Operators
$\hat{b}^{+}$ and $\hat{b}$ satisfy the boson commutation rules:
\begin{eqnarray}\label{komut b}
[\hat{b}_{\vkr l},\hat{b}_{\vec{k'}_{\parallel}l'}^{+}] =
\delta_{\vkr \vec{k'}_{\parallel}}\delta_{ll'}, & [\hat{b}_{\vkr
l},\hat{b}_{\vec{k'}_{\parallel}l'}]=0.&
\end{eqnarray}

By transformation (\ref{op a+a}) Hamiltonian (\ref{Ham
bil}) becomes:
\begin{equation}\label{Ham Four}
\widehat{\cal H}= \sum_{\vkr ,ll'}\left(P_{ll'}(\vkr )
\hat{b}_{\vkr l}^{+}\hat{b}_{\vkr l'} +\frac{1}{2}Q_{ll'}(\vkr )
\hat{b}_{\vkr l}\hat{b}_{-\vkr l'}
+\frac{1}{2}Q_{ll'}^{\ast}(\vkr ) \hat{b}_{\vkr l}^{+}
\hat{b}_{-\vkr l'}^{+}\right) ,
\end{equation}
where:
\begin{eqnarray}\label{Plk}
P_{ll'}(\vkr ) &=& -2\sqrt{S_{l}S_{l'}}J_{ll'}
A_{l}^{\ast}\cdot A_{l'} \Gamma_{ll'}^{\vkr } \\
 & & +\delta_{ll'}\left[2\sum_{n}z_{ln}J_{ln}S_{n}
\vec{\gamma}_{l}\cdot\vec{\gamma}_{n}+
g\mu_{B}\vec{H}^{eff}_{l}\cdot\vec{\gamma}_{l}-D_{l}\left(S_{l}
-\frac{1}{2}\right)\left(1-3\cos^{2}\theta\right)\right] ,
\nonumber
\end{eqnarray}
\begin{equation}\label{Qlk}
Q_{ll'}(\vkr )= -2\sqrt{S_{l}S_{l'}}J_{ll'}A_{l}\cdot A_{l'}
\left(\Gamma_{ll'}^{\vkr }\right)^{\ast} -\delta_{ll'}
\left[D_{l}\sqrt{S_{l}\left(S_{l}-\frac{1}{2}\right)}
\left(1-\cos^{2}\theta\right)\right].
\end{equation}
Term $\Gamma_{ll'}^{\vkr }$, referred to as {\it structural
sum}, is defined as the following sum over spin neighbours:
\begin{equation}\label{Gamma def}
\Gamma_{ll'}^{\pm\vkr }=\sum_{\vec{j}'}e^{\pm i \vkr
\cdot(\vec{j}-\vec{j}')},\qquad (\vec{j}\in l, \: \vec{j}'\in l'),
\end{equation}
and satisfies the following relations:
\begin{eqnarray}\label{Gamma wlasn}
\Gamma_{ll'}^{\ast\vkr } = \Gamma_{ll'}^{-\vkr } , &
\Gamma_{l'l}^{\vkr } = \Gamma_{ll'}^{\ast\vkr } , &
\Gamma_{ll'}^{0}=z_{ll'} .
\end{eqnarray}

\subsection{Transformation along the surface normal}

\hspace*{\parindent} In the last step of the diagonalization
procedure, canonical Tyablikov-Bogolyubov transformation
\cite{Tiablikow} (along the film surface normal) shall be applied
to Hamiltonian (\ref{Ham Four}):
\begin{eqnarray}\label{op b+b}
&& \hat{b}_{\vkr l}= \sum_{\kp}\left[ u_{l}(\kp)
\hat{\xi}_{\vkr\kp} + v_{l}^{\ast}(-\kp) \hat{\xi}_{-\vkr
,-\kp}^{+} \right] , \\ \nonumber && \hat{b}_{-\vkr l}^{+}=
\sum_{\kp}\left[ u_{l}^{\ast}(-\kp)\hat{\xi}_{-\vkr ,-\kp}^{+} +
v_{l}(\kp)\hat{\xi}_{\vkr \kp}\right] ,
\end{eqnarray}
where $\hat{\xi}^{+}_{\vkr \kp}$ and $\hat{\xi}_{\vkr \kp}$ are
operators of creation and annihilation, respectively, of a spin
wave with energy $E(\vkr ,\kp)$ and wave vector $\vec{k}=[\vkr
,\kp]$, components $\vkr$ and $\kp$ being, respectively, parallel
and perpendicular to the surface. When expressed by operators
$\hat{\xi}_{\vkr \kp}$ and $\hat{\xi}^{+}_{\vkr \kp}$, the
Hamiltonian becomes diagonal:
\begin{equation}\label{Ham diag}
\widehat{\cal H}= \sum_{\vkr ,\kp} E(\vkr ,\kp) \hat{\xi}_{\vkr
\kp}^{+} \hat{\xi}_{\vkr \kp}+const.
\end{equation}
Hamiltonian (\ref{Ham diag}) is, by assumption, Hermitian, which
implies that $E(\vkr , \kp) = E^{\ast}(\vkr , \kp)$. For
transformation (\ref{op b+b}) to result in the diagonal form of
the Hamiltonian, functions $u_{l}(\kp)$ and $v_{l}(\kp)$ must
satisfy the following conditions of orthonormality \cite{HP94}:
  \begin{eqnarray}\label{uv war ort}
   &  & \sum_{l}\left[u_{l}(\kp)u_{l}^{\ast}(\kp ') -
v_{l}(\kp)v_{l}^{\ast}(\kp ')\right] = \delta_{\kp\kp '} , \\
\nonumber
   &  & \sum_{l}\left[u_{l}(\kp)v_{l}(-\kp) -
v_{l}(\kp)u_{l}(-\kp)\right] = 0 , \\ \nonumber
   &  & \sum_{\kp}\left[u_{l}(\kp)u_{l'}^{\ast}(\kp) -
v_{l'}(-\kp)v_{l}^{\ast}(-\kp)\right] = \delta_{ll'} , \\
   &  & \sum_{\kp}\left[u_{l}(\kp)v_{l'}^{\ast}(\kp) -
v_{l}^{\ast}(-\kp)u_{l'}(-\kp)\right] = 0 , \nonumber
  \end{eqnarray}
and, moreover, they must be solutions of so-called {\it
Tyablikov-Bogolyubov equations}.

In order to find functions $u_{l}(\kp)$ and $v_{l}(\kp)$ in their
explicit forms, we shall write the Heisenberg equations of motion
for operators $\hat{b}_{\vkr l}$, $\hat{\xi}_{\vkr \kp}$ and
$\hat{\xi}^{+}_{\vkr \kp}$:
\begin{equation}\label{ruch b}
i\dot{\hat{b}_{\vkr l}}= \left[ \hat{b}_{\vkr l},\widehat{\cal
H}\right]= \sum_{l'} \left[P_{ll'}(\vkr )\hat{b}_{\vkr l'}+
Q^{\ast}_{ll'}(\vkr ) \hat{b}^{+}_{-\vkr l'}\right] ,
\end{equation}
\begin{eqnarray}\label{ruch xi+xi}
&& i\dot{\hat{\xi}_{\vkr \kp}}= \left[\hat{\xi}_{\vkr
\kp},\widehat{\cal H}\right]= E(\vkr ,\kp)\hat{\xi_{\vkr \kp}} ,
\\ \nonumber
&& i\dot{\hat{\xi}_{-\vkr ,-\kp}^{+}}= \left[ \hat{\xi}_{-\vkr
,-\kp}^{+},\widehat{\cal H} \right]= E(\vkr ,\kp)\hat{\xi_{\vkr
\kp}} .
\end{eqnarray}
By inserting (\ref{op b+b}) into (\ref{ruch b}), and using
(\ref{ruch xi+xi}), the following equation is obtained:
\begin{eqnarray*}
\sum_{\kp}\left[u_{l}(\kp)E(\vkr ,\kp) \hat{\xi} _{\vkr \kp} -
v_{l}^{\ast}(-\kp) E(-\vkr ,-\kp)\hat{\xi}^{+}_{-\vkr ,-\kp}
\right] & = & \nonumber \\ \sum_{\kp}\sum_{l'} \left[P_{ll'}(\vkr
) \left(u_{l'}(\kp) \hat{\xi} _{\vkr \kp} + v_{l'}^{\ast}(-\kp)
\hat{\xi}^{+}
_{-\vkr ,-\kp}\right) \right. & + & \nonumber \\
\left. Q_{ll'}^{\ast} (\vkr ) \left(u_{l'}^{\ast}(-\kp)
\hat{\xi}^{+} _{-\vkr ,-\kp} + v_{l'}(\kp) \hat{\xi} _{\vkr
\kp}\right)\right] , &&
\end{eqnarray*}
equivalent to a set of $2L$ equations, if  $\hat{\xi}_{\vkr
\kp}$and $\hat{\xi}^{+}_{-\vkr ,-\kp}$ are linearly independent:
\[
 u_{l}(\kp)E(\vkr ,\kp)= \sum_{l'}\left[
P_{ll'}(\vkr )u_{l'}(\kp)+ Q^{\ast}_{ll'}(\vkr )v_{l'}(\kp)
\right] ,
\]
\[
-v_{l}^{\ast}(-\kp)E(-\vkr ,-\kp)= \sum_{l'}\left[ P_{ll'}(\vkr
)v_{l'}^{\ast}(-\kp)+ Q_{ll'}^{\ast}(\vkr )u_{l'}^{\ast}(-\kp)
\right] .
\]
Reversing the direction of $\vec{k}$ in the second equation
($\vkr\rightarrow -\vkr$, $\kp \rightarrow -\kp$) finally leads
to:
\begin{eqnarray}\label{row TiaBog}
&& u_{l}(\kp)E(\vkr ,\kp)= \sum_{l'}\left[ P_{ll'}(\vkr
)u_{l'}(\kp)+ Q^{\ast}_{ll'}(\vkr )v_{l'}(\kp) \right] , \\
\nonumber && -v_{l}(\kp)E(\vkr ,\kp)= \sum_{l'}\left[
P^{\ast}_{ll'}(-\vkr )v_{l'}(\kp)+ Q_{ll'}(-\vkr )u_{l'}(\kp)
\right] .
\end{eqnarray}
This is the set of Tyablikov-Bogolyubov equations corresponding
to our problem. The spin wave energy, $E(\vkr,\kp)$, as well as
functions $u_{l}(\kp)$ and $v_{l}(\kp)$, describing the spin wave
precession, can be deduced from its solution.

\section{The particular case: ferromagnetic bilayer film}\label{rozdz DW}

\subsection{General form of the bilayer Hamiltonian}\label{rozdz Ham
DW}

\begin{figure}
  \centering
  \includegraphics{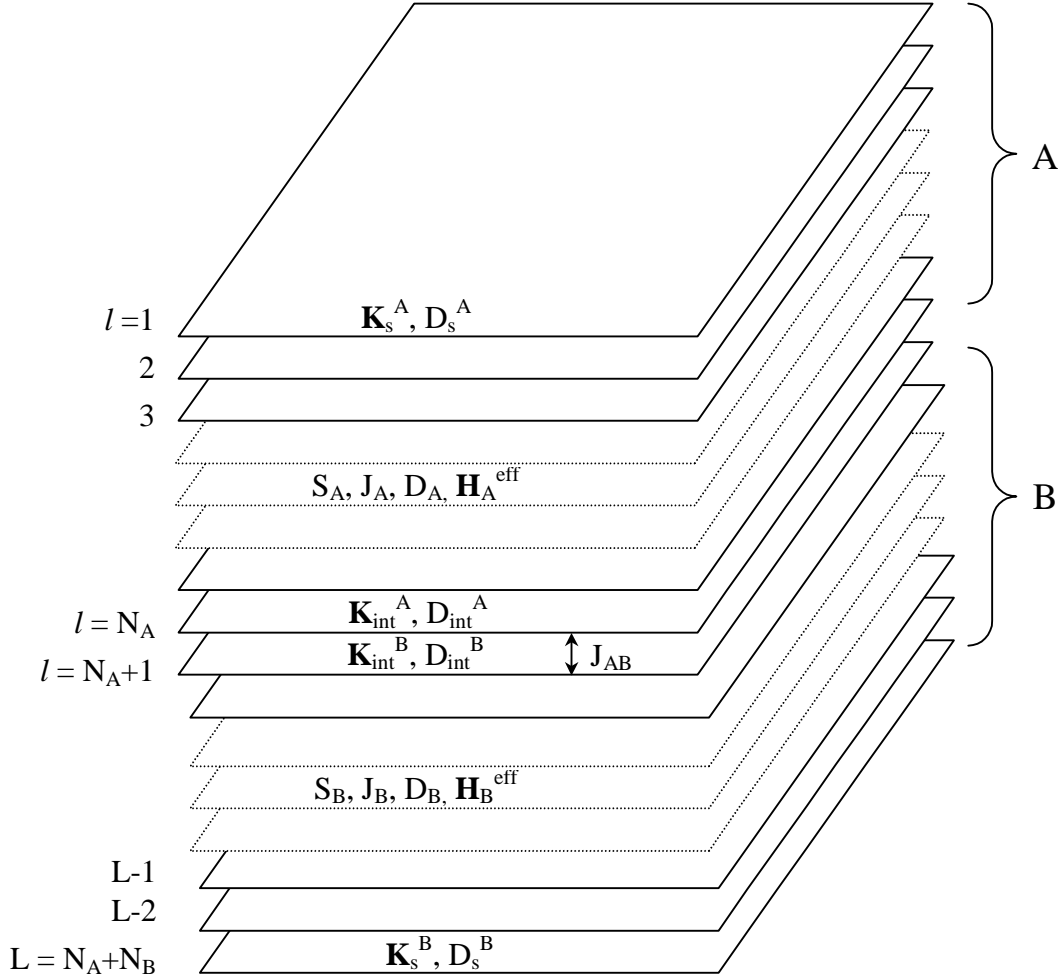}
  %\vspace{15cm}
\caption[]{The planar magnetic bilayer film model (see the text
for detailed description).}\label{rys model DW}
\end{figure}

\hspace*{\parindent} Let us consider now a magnetic bilayer film,
composed of two {\it homogeneous} sublayers A and B having
identical crystallographic structure; each sublayer is assumed to
contain a number of lattice planes, $N_{A}$ and $N_{B}$,
respectively (Fig. \ref{rys model DW}). Moreover, the spins in
the lattice nodes, as well as the exchange interactions between
the nearest neighbours, are assumed to be equal within each
sublayer, though they can differ between the sublayers. The same
rule applies to the other parameters, such as the {\it bulk}
anisotropy constant and the effective field. Besides the standard
notion of {\it bulk} anisotropy, $D_{A(B)}$, two other quantities
shall be used: {\it surface} anisotropy $D^{A(B)}_{s}$ and {\it
interface} anisotropy $D^{A(B)}_{int}$.

% Fig. 2

The exchange interaction through the interface shall be described
by introducing the {\it interface exchange integral}, $J_{AB}$,
positive for ferromagnetic interface coupling, negative for
antiferromagnetic interface coupling, and zero in the case where
no coupling is present between the sublayers. In practice, a thin
non-magnetic interlayer is inserted between the ferromagnetic
thin films A and B, and the interface properties
($D^{A(B)}_{int}$, $J_{AB}$) are determined by the interlayer
thickness and material.

With these assumptions, the matrix form of (\ref{row TiaBog}) is:
\begin{equation}\label{row mac}
\left(\begin{array}{cc}
\begin{array}{cc} X_{A} & X_{AB} \\ X_{AB}^{\dag} & X_{B} \end{array}
 & Y \\ Y &
\begin{array}{cc} X_{A} & X_{AB} \\ X_{AB}^{\dag} & X_{B} \end{array}
\end{array}\right)_{2L\times 2L} \left(\begin{array}{c} U \\ V
\end{array}\right) =E \left(\begin{array}{c} U \\ -V
\end{array}\right) ,
\end{equation}
$U$ and $V$ being defined as follows:
\begin{equation}\label{UV wek}
\begin{array}{cc}
  U=\left[\begin{array}{c}
  U_{1} \\ U_{2} \\ \vdots \\ U_{L_{A}} \\ U_{L_{A}+1} \\
  \vdots \\ U_{L-1} \\ U_{L} \end{array}\right] ,
& V=\left[\begin{array}{c}
  V_{1} \\ V_{2} \\ \vdots \\ V_{L_{A}} \\ V_{L_{A}+1} \\
  \vdots \\ V_{L-1} \\ V_{L} \end{array}\right] .
\end{array}
\end{equation}
In films obtained from cubic crystal surface cuts (001), (110) or
(111), the nearest neighbours of a plane $l$ spin lie in planes
up to $l+3$ \cite{SM98a,SM98b,SM00a,SM00b}, so the general form
of matrices $X$ and $Y$ is seven-diagonal:

\hspace*{-1.2cm}\parbox{16cm}{
\begin{equation}\label{XA mac}
X_{A}=\left[
\begin{array}{ccccccccc}
R_{A}-a_{0} & C_{A} & D_{A} & F_{A} & & & & & \\
C_{A}^{\ast} & R_{A}-a_{1} & C_{A} & D_{A} & F_{A} & & & & \\
D_{A}^{\ast} & C_{A}^{\ast} & R_{A}-a_{2} & C_{A} & D_{A} & F_{A} & & & \\
F_{A}^{\ast} & D_{A}^{\ast} & C_{A}^{\ast} & R_{A} & C_{A} & D_{A} & F_{A} & & \\
& \ddots & \ddots & \ddots & \ddots & \ddots & \ddots & \ddots & \\
& & F_{A}^{\ast} & D_{A}^{\ast} & C_{A}^{\ast} & R_{A} & C_{A} & D_{A} & F_{A} \\
& & & F_{A}^{\ast} & D_{A}^{\ast} & C_{A}^{\ast} & R_{A}-b_{2} & C_{A} & D_{A} \\
& & & & F_{A}^{\ast} & D_{A}^{\ast} & C_{A}^{\ast} & R_{A}-b_{1} & C_{A} \\
& & & & & F_{A}^{\ast} & D_{A}^{\ast} & C_{A}^{\ast} & R_{A}-b_{0}
\end{array}
\right]_{L_{A}\times L_{A}},
\end{equation}}

\hspace*{-1.2cm}\parbox{16cm}{
\begin{equation}\label{XB mac}
X_{B}=\left[
\begin{array}{ccccccccc}
R_{B}-c_{0} & C_{B} & D_{B} & F_{B} & & & & & \\
C_{B}^{\ast} & R_{B}-c_{1} & C_{B} & D_{B} & F_{B} & & & & \\
D_{B}^{\ast} & C_{B}^{\ast} & R_{B}-c_{2} & C_{B} & D_{B} & F_{B} & & & \\
F_{B}^{\ast} & D_{B}^{\ast} & C_{B}^{\ast} & R_{B} & C_{B} & D_{B} & F_{B} & & \\
& \ddots & \ddots & \ddots & \ddots & \ddots & \ddots &
\ddots & \\
& & F_{B}^{\ast} & D_{B}^{\ast} & C_{B}^{\ast} & R_{B} & C_{B} & D_{B} & F_{B} \\
& & & F_{B}^{\ast} & D_{B}^{\ast} & C_{B}^{\ast} & R_{B}-d_{2} & C_{B} & D_{B} \\
& & & & F_{B}^{\ast} & D_{B}^{\ast} & C_{B}^{\ast} & R_{B}-d_{1} & C_{B} \\
& & & & & F_{B}^{\ast} & D_{B}^{\ast} & C_{B}^{\ast} & R_{B}-d_{0}
\end{array}
\right]_{L_{B}\times L_{B}},
\end{equation}}

\begin{equation}\label{XAB mac}
X_{AB}=\left[
\begin{array}{ccccc}
\vdots & \vdots & \vdots & \vdots & \\
0 & 0 & 0 & 0 & \ldots \\
F_{AB} & 0 & 0 & 0 & \cdots \\
D_{AB} & F_{AB} & 0 & 0 & \ldots \\
C_{AB} & D_{AB} & F_{AB} & 0 & \ldots \\
\end{array}
\right]_{L_{A}\times L_{B}},
\end{equation}

\hspace*{-1.2cm}\parbox{16cm}{
\begin{equation}\label{Y mac}
Y=\left[
\begin{array}{cccccccccccc}
 R_{A,s}^{y} & & & & & & & & & & \\
 & R_{A}^{y} & & & & & & & & & \\
 & & \ddots & & & & & & & & \\
 & & & R_{A}^{y} & 0 & 0 & F_{AB}^{y} & & & \\
 & & & 0 & R_{A}^{y} & 0 & D_{AB}^{y} & F_{AB}^{y} & & \\
 & & & 0 & 0 & R_{A,int}^{y} & C_{AB}^{y} & D_{AB}^{y} & F_{AB}^{y} & & \\
 & & & (F_{AB}^{y})^{\ast} & (D_{AB}^{y})^{\ast} & (C_{AB}^{y})^{\ast} & R_{B,int}^{y} & 0 & 0 & & \\
 & & & & (F_{AB}^{y})^{\ast} & (D_{AB}^{y})^{\ast} & 0 & R_{B}^{y} & 0 & & \\
 & & & & & (F_{AB}^{y})^{\ast} & 0 & 0 & R_{B}^{y} & & & \\
 & & & & & & & & & \ddots & & \\
 & & & & & & & & & & R_{B}^{y} & \\
 & & & & & & & & & & & R_{B,s}^{y}
\end{array}\right]_{L\times L}.
\end{equation}}

The symbols used above are defined as follows:
\begin{eqnarray}\label{CDF AB}
C_{i}=-2S_{i}J_{i}\Gamma_{1} , & & \nonumber \\
D_{i}=-2S_{i}J_{i}\Gamma_{2} , & & \\
F_{i}=-2S_{i}J_{i}\Gamma_{3} , & & \nonumber
\end{eqnarray}
\begin{eqnarray}\label{R AB}
R_{i} & = & -2S_{i}J_{i}\Gamma_{0} +
g\mu_{B}\vec{H}_{i}^{eff}\cdot\vec{\gamma}_{i} +
D_{i}\left(S_{i}-\frac{1}{2}\right)\left(3\cos^{2}\theta_{i}-1\right)
\nonumber \\
 & & + 2[z_{0}^{i}+2(z_{1}^{i}+z_{2}^{i}+z_{3}^{i})]S_{i}J_{i} ,
\end{eqnarray}
\begin{eqnarray}\label{Ry AB}
R_{i,(int,s)}^{y} & = &
-D_{i,(int,s)}\sqrt{S_{i}\left(S_{i}-\frac{1}{2}\right)}\sin^{2}\theta_{i},
\end{eqnarray}
where $i=A;B$, and:
\begin{eqnarray}\label{CDF int}
C_{AB} & = & -2\sqrt{S_{A}S_{B}}J_{AB}\vec{A}_{A}^{\ast}\cdot
  \vec{A}_{B}\Gamma_{1} , \nonumber \\
D_{AB} & = & -2\sqrt{S_{A}S_{B}}J_{AB}\vec{A}_{A}^{\ast}\cdot
  \vec{A}_{B}\Gamma_{2} , \\
F_{AB} & = & -2\sqrt{S_{A}S_{B}}J_{AB}\vec{A}_{A}^{\ast}\cdot
  \vec{A}_{B}\Gamma_{3} , \nonumber
\end{eqnarray}
\begin{eqnarray}\label{CDFy int}
C_{AB}^{y} & = & -2\sqrt{S_{A}S_{B}}J_{AB}\vec{A}_{A}\cdot
  \vec{A}_{B}\Gamma_{1} , \nonumber \\
D_{AB}^{y} & = & -2\sqrt{S_{A}S_{B}}J_{AB}\vec{A}_{A}\cdot
  \vec{A}_{B}\Gamma_{2} , \\
F_{AB}^{y} & = & -2\sqrt{S_{A}S_{B}}J_{AB}\vec{A}_{A}\cdot
  \vec{A}_{B}\Gamma_{3} , \nonumber
\end{eqnarray}
\begin{eqnarray}\label{abcd}
a_{0} & = & 2(z_{1}^{A}+z_{2}^{A}+z_{3}^{A})S_{A}J_{A} -
  g\mu_{B}\vec{K}_{s}^{A}\cdot\vec{\gamma}_{A} -
  (D_{s}^{A}-D_{A})(S_{A}-\frac{1}{2})(3\cos^{2}\theta_{A}-1) ,
  \nonumber \\
a_{1} & = & 2(z_{2}^{A}+z_{3}^{A})S_{A}J_{A} , \nonumber \\
a_{2} & = & 2z_{3}^{A}S_{A}J_{A} , \nonumber \\
b_{2} & = & 2z_{3}^{A}S_{A}J_{A} - 2z_{3}^{B}S_{B}J_{AB}
  \vec{\gamma}_{A}\cdot\vec{\gamma}_{B} , \nonumber \\
b_{1} & = & 2(z_{2}^{A}+z_{3}^{A})S_{A}J_{A} -
  2(z_{2}^{B}+z_{3}^{B})S_{B}J_{AB}
  \vec{\gamma}_{A}\cdot\vec{\gamma}_{B} , \nonumber \\
b_{0} & = & 2(z_{1}^{A}+z_{2}^{A}+z_{3}^{A})S_{A}J_{A} -
  2(z_{1}^{B}+z_{2}^{B}+z_{3}^{B})S_{B}J_{AB}
  \vec{\gamma}_{A}\cdot\vec{\gamma}_{B} - \nonumber \\
& & g\mu_{B}\vec{K}_{int}^{A}\cdot\vec{\gamma}_{A} -
  (D_{int}^{A}-D_{A})(S_{A}-\frac{1}{2})(3\cos^{2}\theta_{A}-1) , \\
c_{0} & = & 2(z_{1}^{B}+z_{2}^{B}+z_{3}^{B})S_{B}J_{B} -
  2(z_{1}^{A}+z_{2}^{A}+z_{3}^{A})S_{A}J_{AB}
  \vec{\gamma}_{A}\cdot\vec{\gamma}_{B} - \nonumber \\
& & g\mu_{B}\vec{K}_{int}^{B}\cdot\vec{\gamma}_{B} -
  (D_{int}^{B}-D_{B})(S_{B}-\frac{1}{2})(3\cos^{2}\theta_{B}-1)
  , \nonumber \\
c_{1} & = & 2(z_{2}^{B}+z_{3}^{B})S_{B}J_{B} -
  2(z_{2}^{A}+z_{3}^{A})S_{A}J_{AB}
  \vec{\gamma}_{A}\cdot\vec{\gamma}_{B} , \nonumber \\
c_{2} & = & 2z_{3}^{B}S_{B}J_{B} - 2z_{3}^{A}S_{A}J_{AB}
  \vec{\gamma}_{A}\cdot\vec{\gamma}_{B} , \nonumber \\
d_{2} & = & 2z_{3}^{B}S_{B}J_{B} , \nonumber \\
d_{1} & = & 2(z_{2}^{B}+z_{3}^{B})S_{B}J_{B} , \nonumber \\
d_{0} & = & 2(z_{1}^{B}+z_{2}^{B}+z_{3}^{B})S_{B}J_{B} -
  g\mu_{B}\vec{K}_{s}^{B}\cdot\vec{\gamma}_{B} -
  (D_{s}^{B}-D_{B})(S_{B}-\frac{1}{2})(3\cos^{2}\theta_{B}-1).\nonumber
\end{eqnarray}
As assumed previously, the direction of quantization appearing in
the above formulae is constant within a single sublayer, but can
differ between the sublayers (i.e. $\vec{\gamma}_{l} =
\vec{\gamma}_{A(B)}$, which implies $\vec{A}_{l} =
\vec{A}_{A(B)}$).

\subsection{The effect of various surface cuts}

\hspace*{\parindent} We shall henceforth consider a bilayer film
whose both sublayers are made of the same material
($S_{A}=S_{B}\equiv S$ and $J_{A}=J_{B}\equiv J$), possible
asymmetry being due only to interface ($D_{int}^{A} \neq
D_{int}^{B}$, $\vec{K}_{int}^{A} \neq \vec{K}_{int}^{B}$) or
surface ($\vec{K}_{s}^{A} \neq \vec{K}_{s}^{B}$) conditions. In
the simplest case, the nearest neighbours of a plane $l$ spin lie
in either the same or a neighbouring plane ($l'=l, l \pm 1$),
which brings the Hamiltonian matrix to a three-diagonal form
\cite{SM02}. This situation takes place for surface cut (001) in
all three cubic crystal types, and for cut (110) in sc and bcc
crystals. As our study is focused on ferromagnetic resonance, in
which standing spin waves only are observed, we assume $\vkr =0$.

In order to simplify the problem, the bilayer spin precession
shall be henceforth assumed to be circular. The spin precession
ellipticity is allowed for in (\ref{row mac}) through matrix $Y$
(\ref{Y mac}), whose diagonal and out-of-diagonal elements are
defined in (\ref{Ry AB}) and (\ref{CDFy int}), respectively. If
the quantization vector $\vec{\gamma}$ has the same direction in
both sublayers, then $\vec{A}_{A} = \vec{A}_{B}$, and the
out-of-diagonal elements of $Y$ vanish, since product
$\vec{A}_{A} \cdot \vec{A}_{B}$ is zero. The diagonal elements of
$Y$ can be divided into three groups, distinguishing bulk,
surface and interface elements. Their different values are due to
the fact that the ellipticity of spin precession on the surface
and the interface differs from that in the bulk. However, if the
spin precession ellipticity is assumed approximately {\it
homogeneous} throughout the bilayer film, all the diagonal
elements of $Y$ become equal; as shown in \cite{HP79}, this does
not affect the {\it relative} intensities of the resonance lines,
and thus matrix $Y$, having no effect on the resonance spectrum,
can be omitted ({\it i.e.} we can assume $Y \equiv 0$). Hence, in
the circular spin precession approximation, set of equations
(\ref{row mac}) can be separated, and reduced to a simple
eigenvalue problem with matrix $X$ only.

With the above-specified assumptions, for cubic crystal surface
cut (001) equation (\ref{row mac}), divided by $2SJ\Gamma_{1}$,
becomes:

\hspace*{-0.5cm}\parbox{16cm}{
\begin{equation}\label{row mac 001}
 \left[
 \begin{tabular}{cccc@{}c@{}c@{}c@{}cccc}
   $2-A_{surf}^{A}$ & $-1$ & & & & \vline & & & \\
   $-1$ & $2$ & $-1$ & & & \vline & & & & \\
    & $\ddots$ & $\ddots$ & $\ddots$ & & \vline & & & \\
    & & $-1$ & $2$ & $-1$ & \vline & & & \\
    \cline{5-7}
    & & & $-1$ \, \vline & \, $2-A_{int}^{A}$ \, & & $-J_{int}$ & \vline \protect{\hfill} & & \\
    \cline{1-4} \cline{8-11}
    & & & \protect{\hfill} \vline & $-J_{int}$ & & \, $2-A_{int}^{B}$ \, & \vline \, $-1$ & & \\
    \cline{5-7}
    & & & & & \vline & $-1$ & $2$ & $-1$ & \\
    & & & & & \vline & & $\ddots$ & $\ddots$ & $\ddots$\\
    & & & & & \vline & & & $-1$ & $2$ & $-1$ \\
    & & & & & \vline & & & & $-1$ & $2-A_{surf}^{B}$ \\
 \end{tabular}
 \right] U = E'U ,
\end{equation}}
the Hamiltonian matrix elements being defined as follows:
\begin{eqnarray}
 A_{surf}^{A(B)} & = & 1 - \underbrace{\frac{g\mu_{B}}{2SJz_{\perp}}
 (\vec{\gamma}\cdot\vec{K}_{s}^{A(B)})}_{\mbox{\normalsize$a_{s0}$}}-
 \underbrace{\frac{D_{s}^{A(B)}(S-\frac{1}{2})}{2SJz_{\perp}}}_{
 \mbox{\normalsize$a_{s2}$}}(3\cos^{2}\theta-1), \\
 & & \nonumber \\
 A_{int}^{A(B)} & = &  1 - J_{int} -
 \underbrace{\frac{g\mu_{B}}{2SJz_{\perp}}
 (\vec{\gamma}\cdot\vec{K}_{int}^{A(B)})}_{\mbox{\normalsize$a_{i0}$}} -
 \underbrace{\frac{D_{int}^{A(B)}(S-\frac{1}{2})}{2SJz_{\perp}}}_{
 \mbox{\normalsize$a_{i2}$}}(3\cos^{2}\theta-1) , \\
 \nonumber \\
 J_{int} & = & J_{AB}/J .
\end{eqnarray}
Properties of function $\Gamma$ (\ref{Gamma wlasn}) imply that
$\Gamma_{ll'}(\vkr=0)=z_{ll'}$, and thus
\begin{equation}
 E'=E/(2SJz_{\perp}) ,
\end{equation}
where $z_{\perp} \equiv z_{l,l+1}$.

Considering symmetric surface or interface conditions, we shall
henceforth omit indices A and B in the respective sublayer
parameter symbols, {\it e.g.} $a_{i2}^{A} = a_{i2}^{B} \equiv
a_{i2}$, or $A_{surf}^{A} = A_{surf}^{B} \equiv A_{surf}$.

Equation (\ref{row mac 001}) provides the basis for our further
investigation. The Hamiltonian matrix corresponds to a simplified
bilayer film model on which a spin-wave spectrum can be studied
as a function of three essential structural magnetic parameters:
the surface parameter, $A_{surf}$, the interface parameter,
$A_{int}$, and the interface coupling, $J_{int}$. Below,
eigenvalue problem (\ref{row mac 001}) shall be solved
numerically only, assuming the spin value equal to one (S=1).

\section{Ambiguity in existing interpretation of bilayer FMR spectra}\label{rozdz dwa piki}

% Fig. 3
\begin{figure}
  \centering
  \includegraphics{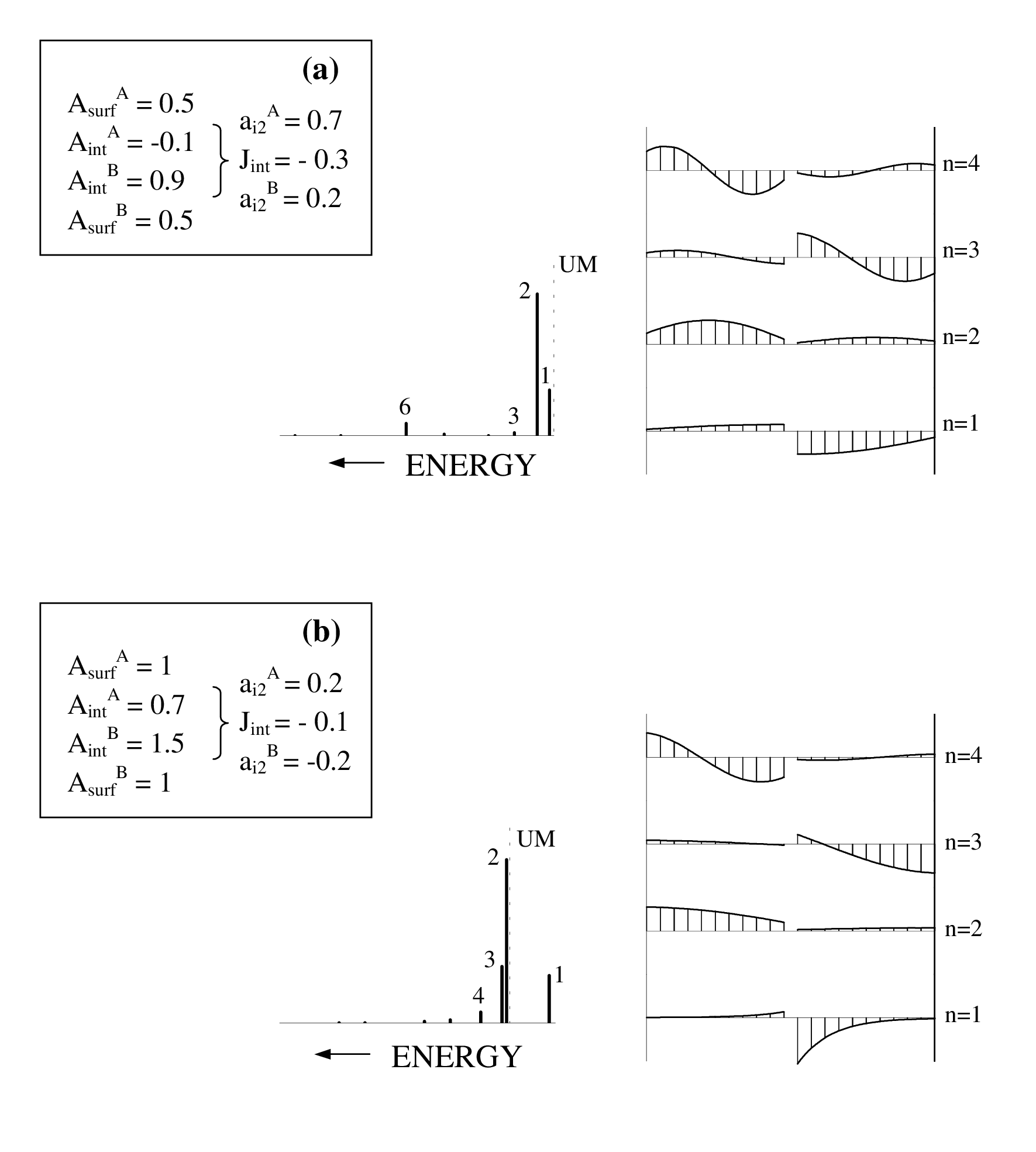}
  %\vspace{18cm}
\caption[]{Profiles of the first ({\it i.e.} having the lowest
energies) bilayer modes and the corresponding resonance spectra
obtained for the parameter values specified in boxes (the values
of interface parameter $A_{int}$ correspond to those assumed for
$J_{int}$ and $a_{i2}$, specified on the right of the bracket).
The dashed line indicates the position of a hypothetic uniform
mode (UM) with $\vec{k} \equiv 0$.
  \protect{\newline}
(a) Asymmetric interface, antiferromagnetic interface coupling.
  \protect{\newline}
(b) Asymmetric interface, interface coupling still
antiferromagnetic, but weaker than in (a). Note that the value of
parameter $a_{i2}^{B}$ is negative.}\label{interp1}
\end{figure}

\addtocounter{figure}{-1}

\begin{figure}
  \centering
    \includegraphics{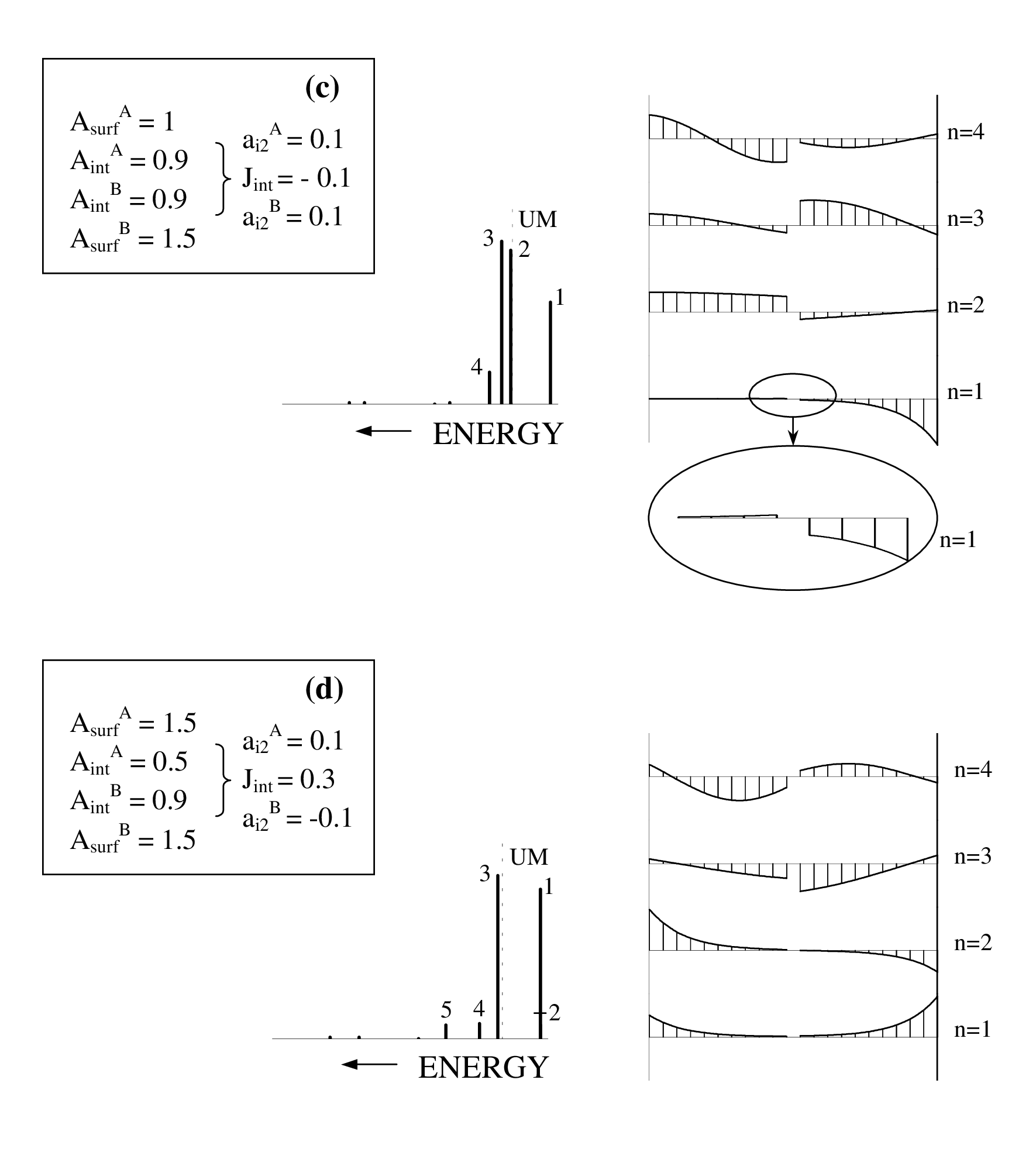}
  \caption[]{c, d
  \protect{\newline}
(c) Asymmetric surfaces, antiferromagnetic interface coupling.
  \protect{\newline}
(d) Asymmetric interface, ferromagnetic interface coupling.}
\end{figure}

\addtocounter{figure}{-1}

\begin{figure}
  \centering
    \includegraphics{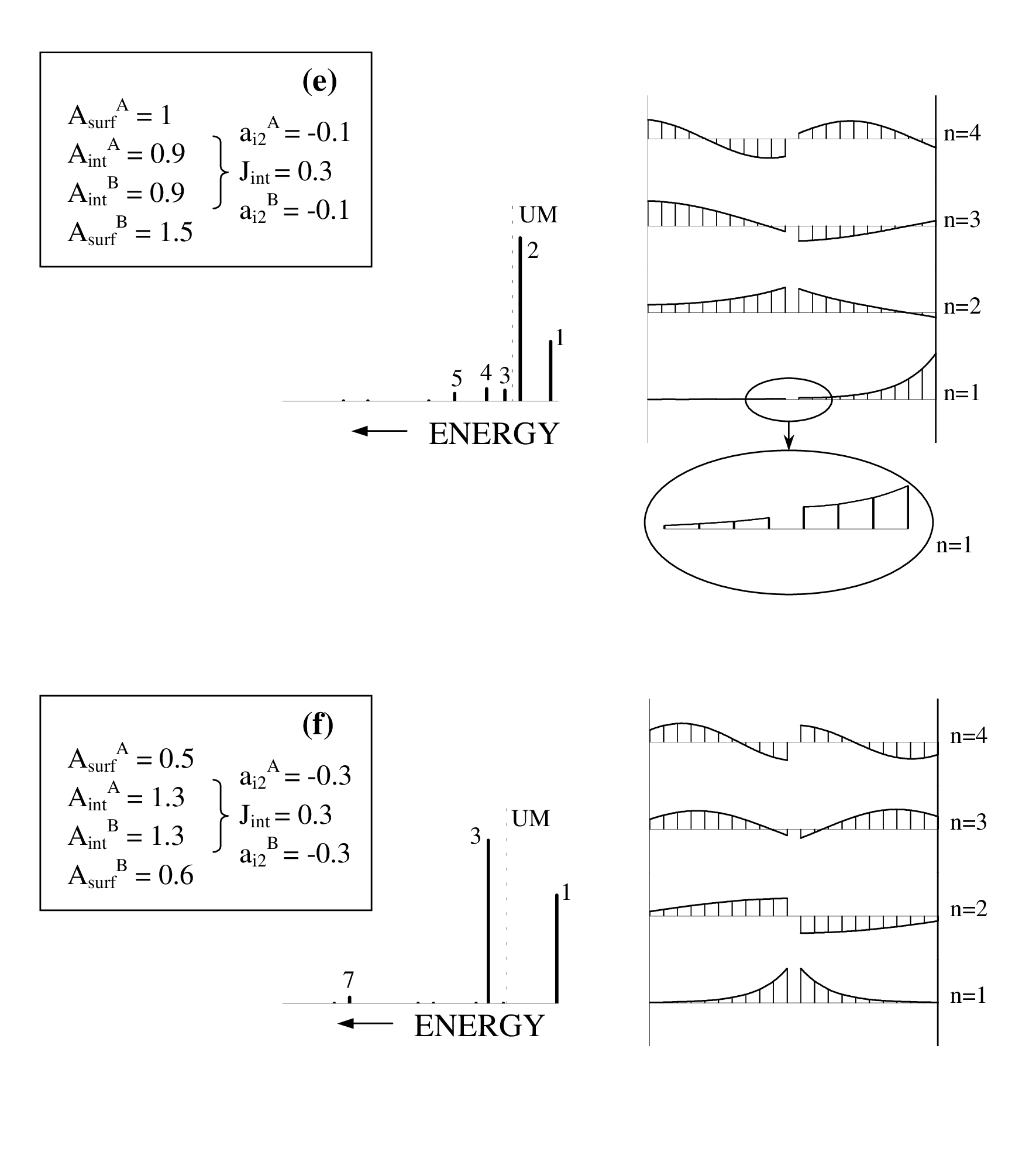}
  \caption[]{e, f
  \protect{\newline}
(e) Asymmetric surfaces, ferromagnetic interface coupling.
  \protect{\newline}
(f) Asymmetric surfaces, ferromagnetic interface coupling.}
\end{figure}

\hspace*{\parindent} In the 'reversed' double-peak FMR spectrum,
reported in bilayer films, the high-intensity line is commonly
interpreted as corresponding to an acoustic mode, the
low-intensity line being related to an optic mode
\cite{Heinrich91,Jin95,Inomata96,Tong99,Lindner01}. This implies
that such 'reversed' spectrum should appear only in bilayers with
{\it antiferromagnetic} interface coupling. Fig. \ref{interp1}
shows examples of bilayer resonance spectra with inverse
intensity arrangement, resulting from our numerical computations;
(a)-(c) were obtained assuming antiferromagnetic interface
coupling \footnote{ This case ({\it i.e.} a bilayer with
antiferromagnetic interface coupling) was also considered in our
previous paper \cite{HP94}. However, as we have just recently
realized, the numerical calculations performed there - for this
particular case only - were incorrect due to the faulty
computational program used in that paper. This resulted in
misinterpretation of the energetically lowest mode as a {\it
symmetric} one. Our present {\it correct} numerical calculations
show that, in fact, this mode is of {\it antisymmetric} nature.
This discrepancy leads to quantitatively different resonance
spectra, however, it does not affect the correctness of our main
hypothesis formulated in \cite{HP94} about the permissible {\it
interface-localized} nature of the first resonance mode.}, while
(d)-(f) correspond to bilayers in which the interface coupling is
ferromagnetic. Perpendicular configuration ($\theta = 0$), as
well as the absence of the uni-directional anisotropy
($a_{i0}=0$) were assumed in all cases. The case depicted in Fig.
\ref{interp1}(a) corresponds to the commonly used interpretation:
optic bulk mode $n=1$ has lower intensity than acoustic mode
$n=2$. In case (b), the high-intensity mode ($n=2$) is still
acoustic, but the low-intensity line ($n=1$) corresponds to an
optic {\it interface} mode. Another possibility is shown in Fig.
\ref{interp1}(c): both first modes, $n=1$ and $n=2$, are of
'optic' nature, $n=2$ being an optic bulk mode and $n=1$ being an
{\it optic} surface mode; the low intensity of the latter is due
not only to its optic character, but also to its localization at
the surface (this case has already been studied in
\cite{Jackson97}). If the interface coupling is ferromagnetic,
three possibilities can occur, their examples shown in Figs.
\ref{interp1}(d)-(f). In case (d), the low-intensity peak ($n=1$)
corresponds to an acoustic surface mode, while the high-intensity
line ($n=3$) is related to an acoustic bulk mode. In (e), the
low-intensity mode ($n=1$) is acoustic and localized at the
surface, and the high-intensity mode ($n=2$) is acoustic and
localized {\it at the interface}. In (f), the low-intensity mode
($n=1$) is acoustic and localized at the interface, while the
high-intensity peak ($n=3$) corresponds to an acoustic bulk mode.

These results clearly indicate that, from the theoretical point
of view, the commonly used interpretation of the double-peak
resonance spectra, relating the high-intensity line to an
acoustic mode, and the low-intensity line to an optic mode, is
not always legitimate.

\section{The effect of interface coupling}\label{rozdz wplyw}

% Fig. 4
\begin{figure}[h]
  \centering
  \includegraphics{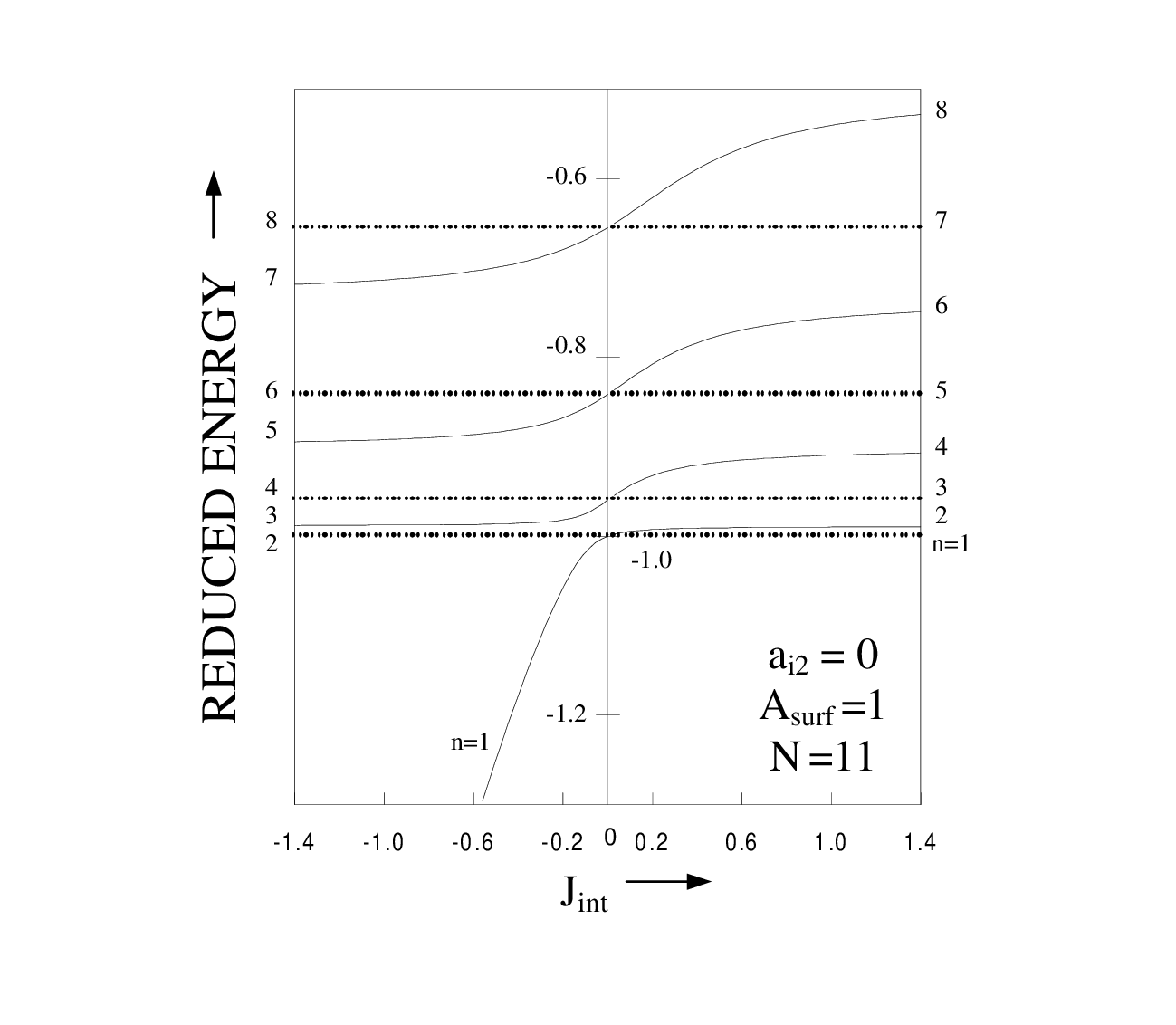}
  %\vspace{10cm}
\caption[]{Energies of symmetric bilayer modes {\it versus} the
interface coupling integral, $J_{int}$. The solid lines
correspond to antisymmetric modes, the dotted lines corresponding
to symmetric ones.}\label{poloz od Jint}
\end{figure}

% Fig. 5
\begin{figure}
  \centering
  \includegraphics{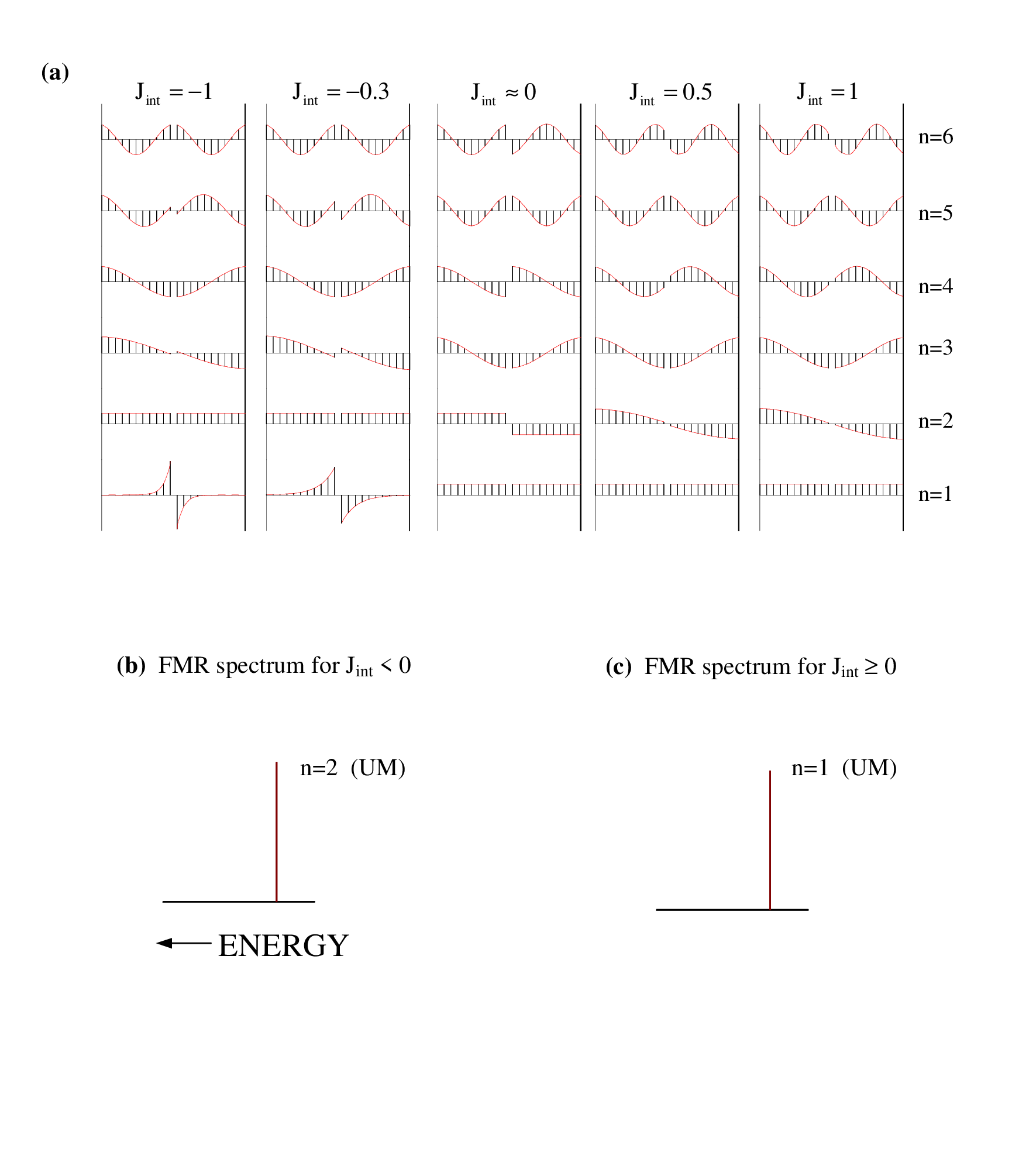}
  %\vspace{20cm}
\caption[]{(a) Profiles of the six lowest modes and (b, c) the
corresponding FMR spectra in a symmetric bilayer, computed
assuming no interface anisotropy ($a_{i2}=0$), the surface spins
having natural freedom ($A_{surf}=1$), for different values of
interface exchange integral $J_{int}$. Each sublayer is composed
of $N = 11$ lattice planes. UM denotes the resonance line
corresponding to the uniform mode excitation.}\label{teor wid 1}
\end{figure}

\hspace*{\parindent} In this Section we shall analyse
ferromagnetic resonance spectra obtained in the so-called
perpendicular configuration, {\it i.e.} for $\theta =0$. For
simplicity reasons, we shall assume that there is no
uni-directional anisotropy on the interface ($a_{i0}=0$), and the
surface spins have 'natural' freedom ($A_{surf} \equiv 1$), Thus,
the only source of anisotropy is the interface {\it uni-axial}
anisotropy. Fig. \ref{poloz od Jint} shows bilayer mode energies
as functions of the interface coupling integral in the absence of
any interface anisotropy ($a_{i2}=0$) (this situation was studied
in \cite{HP94}). If the coupling is ferromagnetic, $J_{int}$ has
no effect on the energy of the odd modes, while in the case of
antiferromagnetic coupling, the modes 'insensitive' to $J_{int}$
variations are even. This result fully corresponds to that
reported in \cite{HP94} and \cite{Castillo98}. An insight into
these functions is provided by the mode profiles analysed versus
the interface coupling integral. Fig. \ref{teor wid 1}(a) shows
profiles of the six lowest modes for five different $J_{int}$
values. In the case of ferromagnetic coupling the interface
coupling integral is found to have little effect on the shape of
these profiles. Mode $n=1$ is always a uniform mode; the even
modes are antisymmetric (and as such do not appear in the
resonance spectrum), and all the odd modes except $n=1$, though
symmetric in the bilayer, are antisymmetric within each sublayer
and thus do not appear in the resonance spectrum either. Hence,
it is only the uniform mode that is observed in the resonance
spectrum (see Fig. \ref{teor wid 1}(c)). When $J_{int}<0$, the
lowest mode is localized at the interface (the localization
becoming stronger as $J_{int}$ absolute value increases), while
mode $n=2$ is uniform. However, also in this case only one
resonance line is observed ($n=2$, see Fig. \ref{teor wid 1}(b)),
as all the odd modes, including the interface mode, are
antisymmetric, and thus do not appear in the resonance spectrum.
Hence, the resonance spectrum is found to be insensitive to the
interface coupling integral variations (since $J_{int}$ has no
effect on the symmetric modes, and affects only the antisymmetric
ones).

% Fig. 6
\begin{figure}
  \centering
  \includegraphics{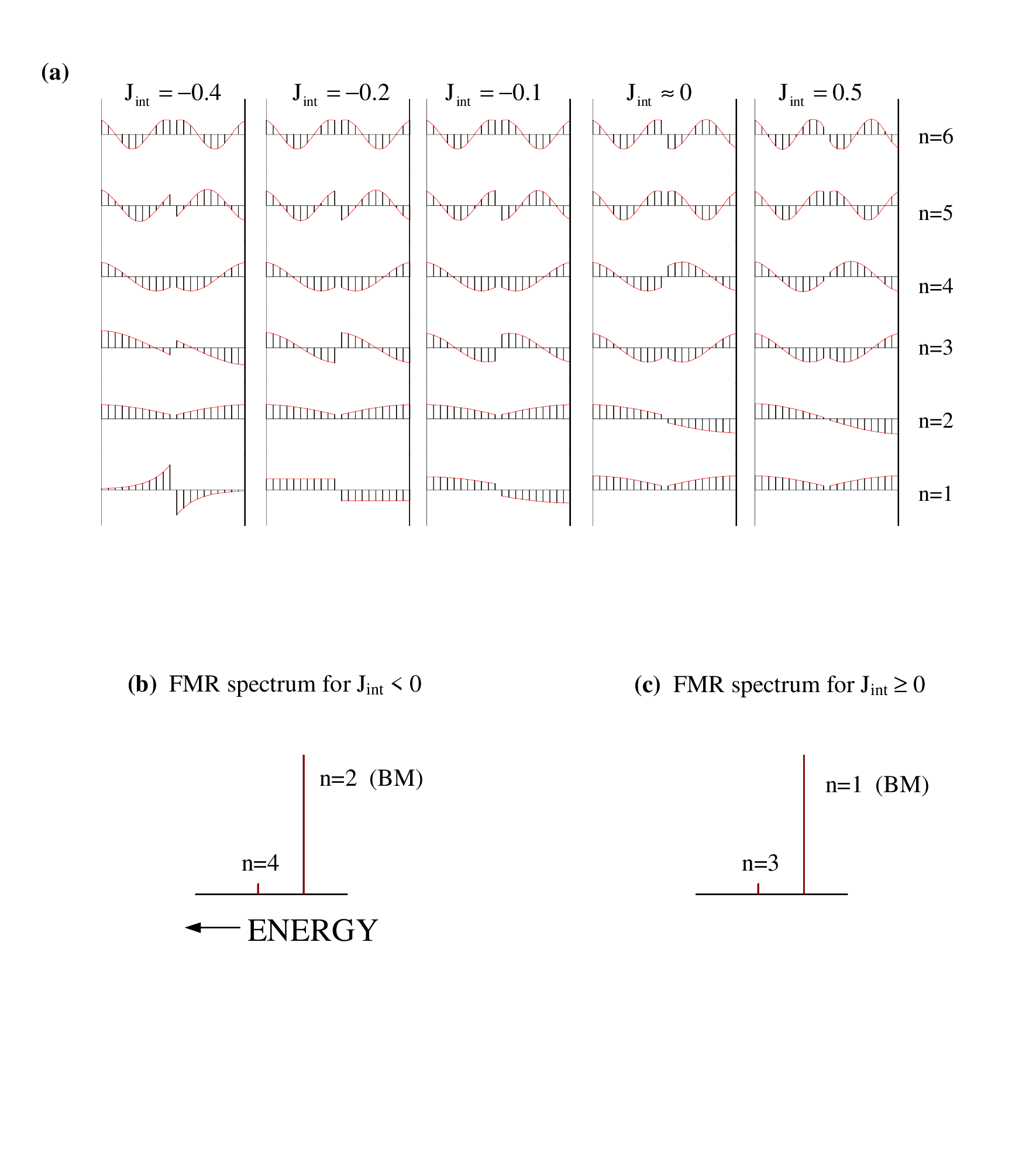}
  %\vspace{20cm}
\caption[]{(a) Profiles of the six lowest modes and (b, c) the
corresponding FMR spectra in a symmetric bilayer, computed
assuming interface anisotropy $a_{i2}=0.2$, the surface spins
having natural freedom ($A_{surf}=1$), for different values of
interface exchange integral $J_{int}$. Each sublayer is composed
of $N = 11$ lattice planes. BM denotes a resonance line
corresponding to the symmetric bulk mode excitation.}\label{teor
wid 2}
\end{figure}

% Fig. 7
\begin{figure}
  \centering
  \includegraphics{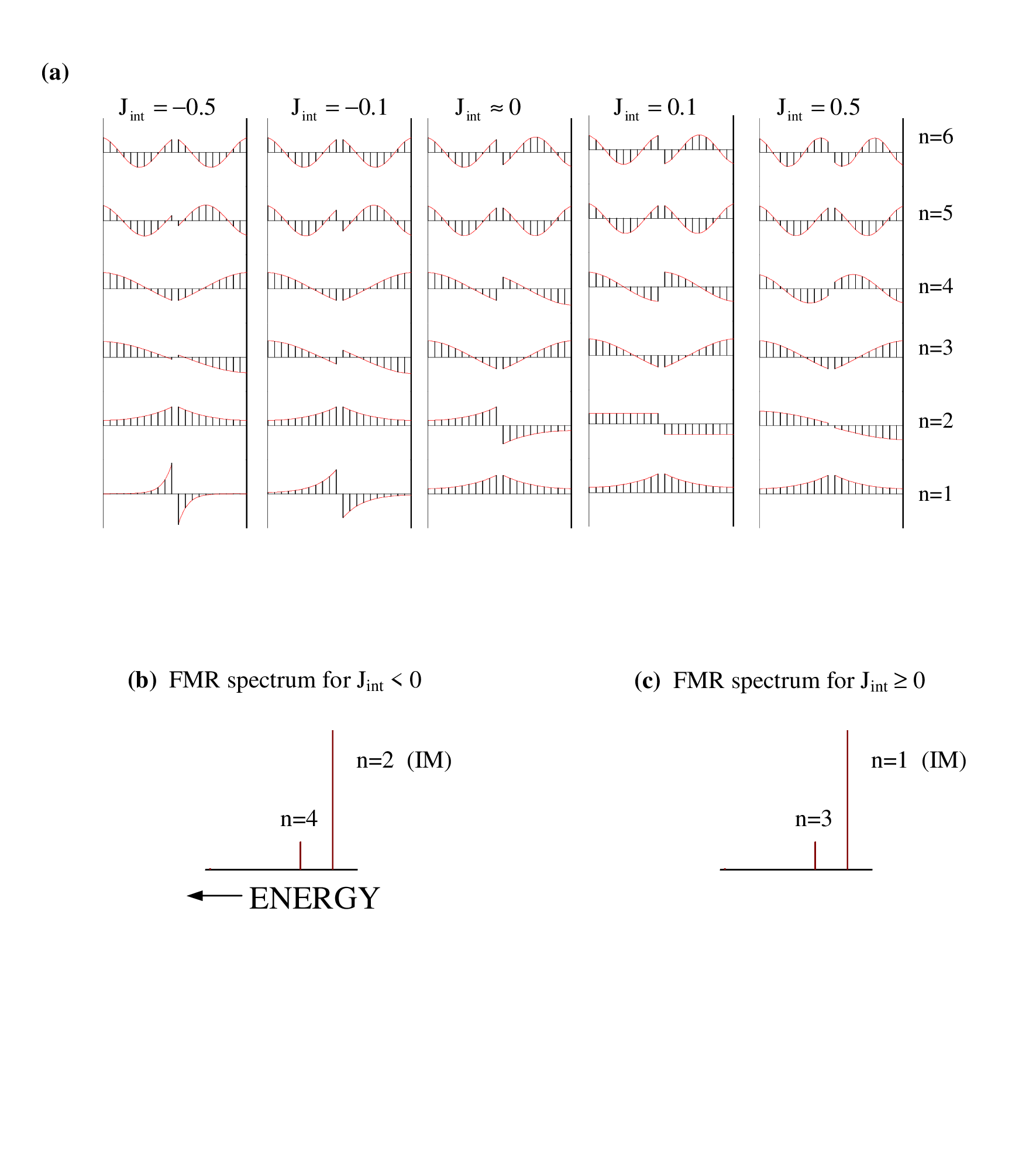}
  %\vspace{20cm}
\caption[]{(a) Profiles of the six lowest modes and (b, c) the
corresponding FMR spectra in a symmetric bilayer, computed
assuming interface anisotropy $a_{i2}=-0.1$, the surface spins
having natural freedom ($A_{surf}=1$), for different values of
interface exchange integral $J_{int}$. Each sublayer is composed
of $N = 11$ lattice planes. IM denotes the resonance line
corresponding to the symmetric interface mode
excitation.}\label{teor wid 3}
\end{figure}

The above conclusion can be equally deduced in a different
reasoning, based on the notion of {\it effective interface
parameter}, $A_{eff}$, introduced in \cite{HP94}. In the
considered perpendicular configuration ($\theta = 0$) this
parameter is expressed as follows:
\begin{equation}\label{param eff int}
A_{eff} = \left\{
\begin{tabular}{ll}
  $ A_{eff}^{s} = 1 - 2a_{i2}$, & for symmetric modes,\\
  $ A_{eff}^{a} = 1 - 2a_{i2} - 2J_{int}$, & for antisymmetric modes.
\end{tabular} \right.
\end{equation}
The above formulae directly indicate that only the antisymmetric
modes depend on $J_{int}$, the symmetric modes being unrelated to
it. This remains valid also when $a_{i2} \neq 0$, as shown in
Figs. \ref{teor wid 2} and \ref{teor wid 3}. However, the FMR
spectrum depicted there contains more than one peak, the
resonance lines corresponding exclusively to either odd or even
modes, for $J_{int}>0$ or $J_{int}<0$, respectively. In each
case, as previously, the FMR spectrum does not depend on $J_{int}$
value.

\section{Critical angle in bilayer resonance spectrum}\label{rozdz
kryt}

\hspace*{\parindent} The resonance spectra considered in the
preceding sections were obtained for the so-called perpendicular
configuration, in which the magnetization vector is oriented
along the film surface normal. In this section, we are going to
investigate configuration effects in resonance spectrum, due to
different orientation of the magnetization vector with respect to
the film surface. Our analysis shall be based on the {\it
effective interface parameter}, $A_{eff}$, a concept introduced
in our earlier study \cite{HP94}. In a symmetric bilayer film,
this parameter is expressed as follows:
\begin{equation}\label{param eff int teta}
A_{eff} = \left\{
\begin{tabular}{ll}
  $ A_{eff}^{s} = 1 - a_{i0} - a_{i2}(3\cos^{2}\theta-1) $ ,
    & for symmetric modes, \\
  $ A_{eff}^{a} = 1 - a_{i0} - a_{i2}(3\cos^{2}\theta-1) - 2J_{int} $,
    & for antisymmetric modes.
\end{tabular} \right.
\end{equation}

Bilayer resonance spectra are generally composed of several
resonance lines (so-called {\it spin wave resonance} - SWR),
though in certain conditions the spectrum reduces to a single
peak. This occurs, for example, when angle $\theta$ between the
magnetization vector and the surface normal takes a particular
value, referred to as {\it critical angle}. The existence of the
critical angle is due to the fact that the effective interface
parameter (\ref{param eff int teta}) is a function of $\theta$.
For simplicity reasons, in our investigation of the critical
angle effect we shall assume that the surface parameter,
$A_{surf}$, does not depend on $\theta$, and that $A_{surf}=1$
({\it i.e.} the spins have 'natural' freedom on both surfaces).

\subsection{The effect of uni-axial anisotropy}

% Fig. 8
\begin{figure}
  \centering
  \includegraphics{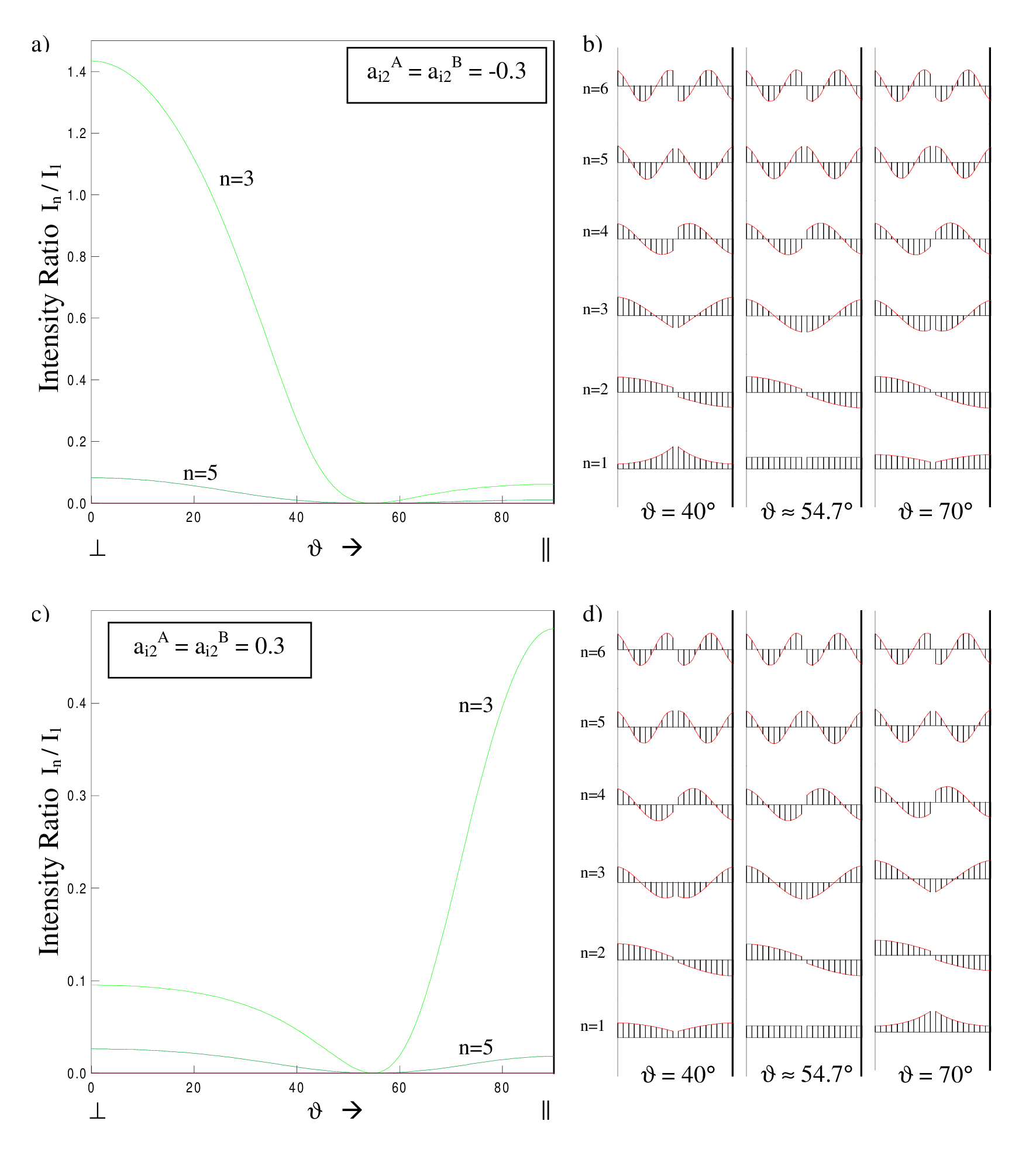}
  %\vspace{20cm}
\caption[]{Resonance line relative intensities {\it versus} the
configuration angle in a bilayer film with symmetric interface
boundary conditions and ferromagnetic coupling  ($J_{int}=0.3$).
The other parameter values assumed: $N=11$, $a_{i0}=0$,
$A_{surf}=1$. Graphs (a) and (c) show intensity ratio
$I_{n}/I_{1}$ plotted {\it versus} angle $\theta$; the
corresponding profiles of the six lowest modes are shown in (b)
and (d) for three different $\theta$ values.}\label{kryt1}
\end{figure}

\hspace*{\parindent} In the first case to be considered, the
effect of the uni-directional anisotropy shall be neglected
($a_{i0}=0$). Figs. \ref{kryt1}(a) and \ref{kryt1}(c) show the
intensities of mode excitations (in relation to that of the first
symmetric mode, $n=1$) as functions of angle $\theta$ in a
symmetric bilayer film (composed of 22 atomic planes, 11 planes
in each sublayer) with ferromagnetic interface coupling
($J_{int}=0.3$). Graphs (a) and (c) were plotted assuming
$a_{i2}^{A}=a_{i2}^{B}=-0.3$, and $a_{i2}^{A}=a_{i2}^{B}=0.3$,
respectively. The corresponding profiles of the six lowest modes
are depicted in Figs. \ref{kryt1}(b) and \ref{kryt1}(d), for
three different values of angle $\theta$: $40^{o}$ ($\theta <
\theta_{crit}$), $54.7^{o}$ ($\theta = \theta_{crit}$) and
$70^{o}$ ($\theta > \theta_{crit}$). (Note that in both cases the
even modes, being antisymmetric, do not appear in the resonance
spectrum).

As shown in Fig. \ref{kryt1}(b), for
$a_{i2}^{A}=a_{i2}^{B}=-0.3$, mode $n=1$, localized at the
interface when $\theta < \theta_{crit}$, becomes a uniform mode
for $\theta = \theta_{crit}$, and a bulk mode when $\theta >
\theta_{crit}$. The other modes are of bulk nature in all $\theta$
range. The resonance intensity of the first bulk mode ($n=3$),
surpassing that of the interface mode at $\theta = 0^{o}$ (see
Fig. \ref{kryt1}(a)), decreases with growing $\theta$ and equals
the interface mode intensity at $\theta \approx 23.4^{o}$. As
$\theta$ continues to increase, the intensities of the bulk modes
decrease still further to vanish completely at $\theta =
\theta_{crit}(\approx 54.7^{o})$, when mode $n=1$ becomes
uniform, and the only one to appear in the SWR spectrum. When
$\theta > \theta_{crit}$, all the modes are of bulk character,
the intensities of modes $n \geq 3$ being much lower than that of
mode $n=1$. Thus, the critical angle is found to separate two
regions in which the relative mode intensities are radically
different.

For $a_{i2}^{A}=a_{i2}^{B}=0.3$ the situation is reversed (Figs.
\ref{kryt1}(c) and \ref{kryt1}(d)). When $\theta <
\theta_{crit}$, all the modes are of bulk character; mode $n=1$
becomes uniform at $\theta = \theta_{crit}$ (the intensities of
all other modes being zero), and localized at the interface when
$\theta > \theta_{crit}$, its intensity remaining the highest in
all $\theta$ range. This case is a 'mirror image' of that
depicted in Figs. \ref{kryt1}(a) and \ref{kryt1}(b); also here,
the critical angle separates two regions in which the relative
mode intensities are completely different.

% Fig. 9
\begin{figure}
  \centering
  \includegraphics{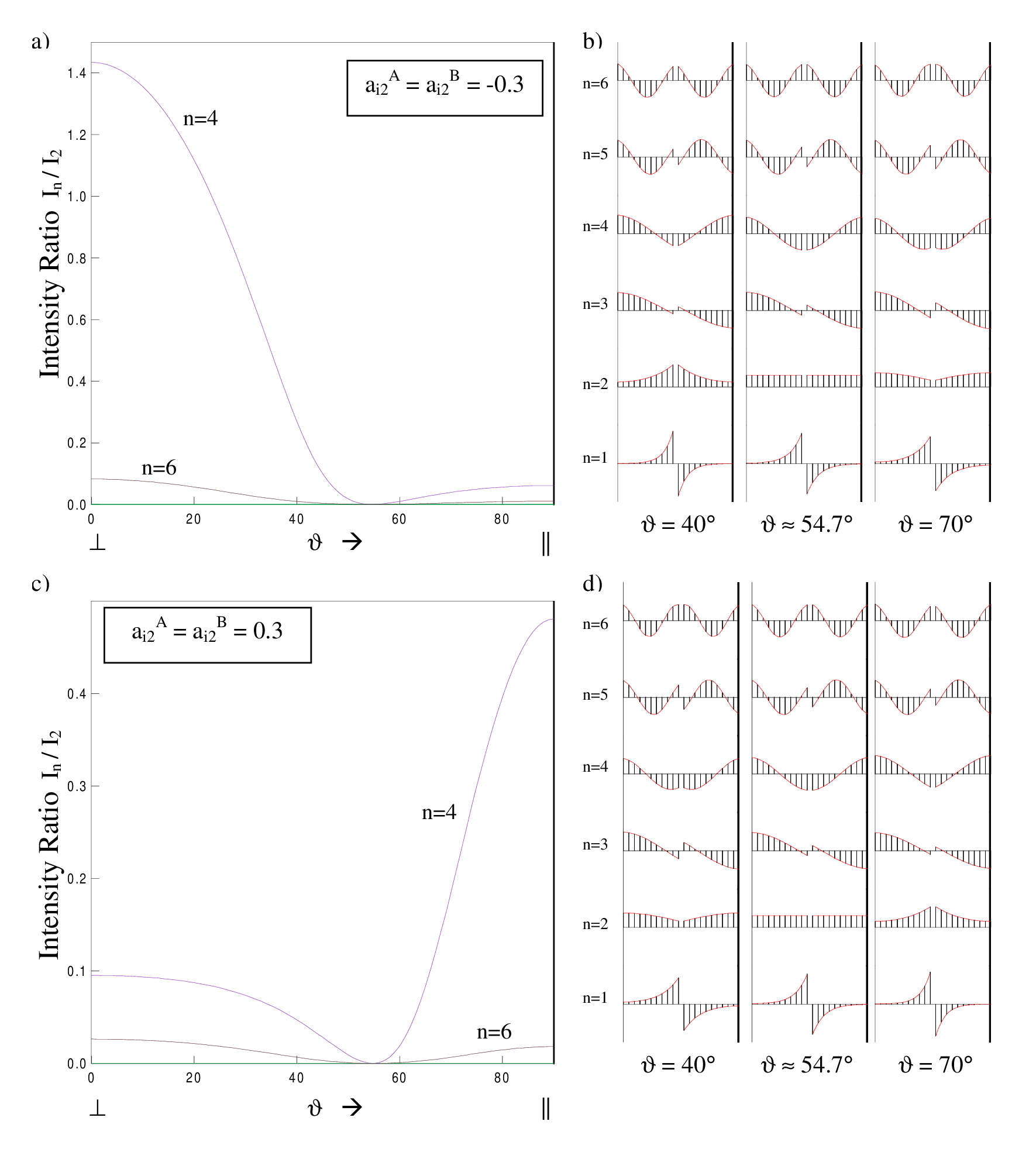}
  %\vspace{20cm}
\caption[]{Resonance line relative intensities {\it versus} the
configuration angle in a bilayer film with symmetric interface
boundary conditions and antiferromagnetic coupling
($J_{int}=-0.3$). The other parameter values assumed: $N=11$,
$a_{i0}=0$, $A_{surf}=1$. Graphs (a) and (c) show intensity ratio
$I_{n}/I_{2}$ plotted {\it versus} angle $\theta$; the
corresponding profiles of the six lowest modes are shown in (b)
and (d) for three different $\theta$ values.}\label{kryt3}
\end{figure}

When the interface coupling becomes antiferromagnetic
($J_{int}=-0.3$, see Fig. \ref{kryt3}), the angle relations of
the relative mode intensities (with respect to the intensity of
the first symmetric mode, $n = 2$) are in principle similar to
those obtained in the case of ferromagnetic coupling, the
critical angle remaining $54.7^{o}$.

From the numerical analysis presented above we deduce that in the
case of symmetric bilayer with surface spins having natural
freedom, the critical angle value is $54.7^{o}$ and does not
depend on either the interface coupling, $J_{int}$, or the
uni-axial anisotropy, $a_{i2}$. Its independence of the interface
coupling is a consequence of the fact that $J_{int}$, having no
effect on the symmetric modes, does not affect the bilayer
resonance spectrum, as shown in Section \ref{rozdz wplyw}. As
regards the uni-axial anisotropy, its variations, though
significantly modifying the SWR spectrum, do not change the
$\theta_{crit}$ value, as the critical angle condition,
$A_{eff}^{s}=1$, is satisfied when $\cos^{2}\theta - 1 = 0$ (or
$\theta = 54.7^{o}$) (\ref{param eff int teta}), $a_{i2}$ being
uninvolved.

% Fig. 10
\begin{figure}
  \centering
  \includegraphics{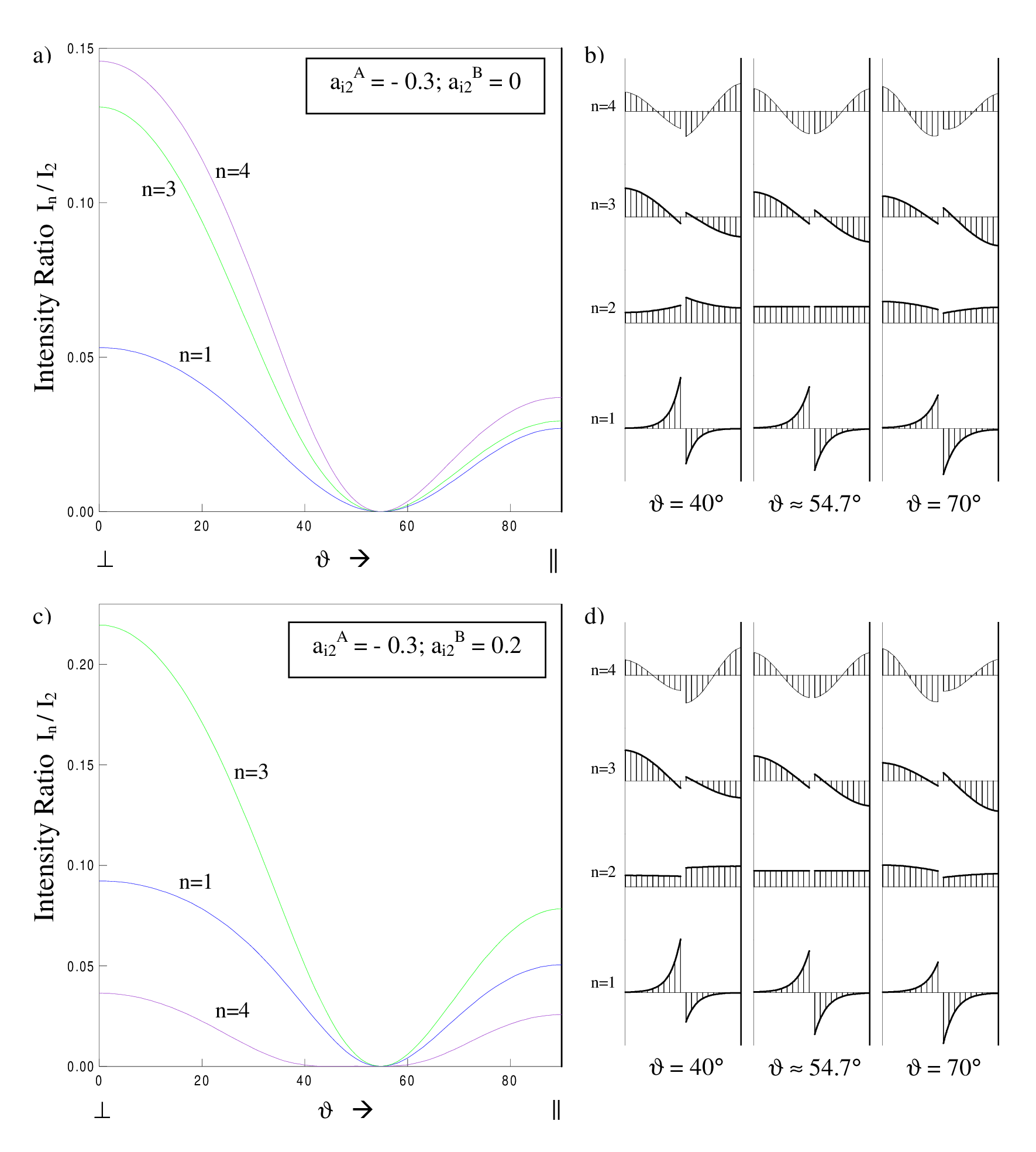}
  %\vspace{20cm}
\caption[]{Resonance line relative intensities {\it versus} the
configuration angle in a bilayer film with asymmetric interface
boundary conditions and antiferromagnetic coupling ($a_{i2}^{A}
\neq a_{i2}^{B}$, $J_{int}=-0.3$). The other parameter values
assumed: $N=11$, $a_{i0}=0$, $A_{surf}=1$. Graphs (a) and (c)
show intensity ratio $I_{n}/I_{2}$ plotted {\it versus} angle
$\theta$ for two different pairs $a_{i2}^{A}$, $a_{i2}^{B}$; the
corresponding profiles of the six lowest modes are shown in (b)
and (d) for three different $\theta$ values.}\label{kryt4}
\end{figure}

Let us now consider a bilayer film with asymmetric interface
conditions. Fig. \ref{kryt4} shows the corresponding relative
mode intensities plotted {\it versus} the configuration angle,
$\theta$. In the considered asymmetric bilayer film, each
sublayer is composed of 11 atomic planes, the interface coupling
is antiferromagnetic ($J_{int}=-0.3$), and the assumed interface
uni-axial anisotropy values in sublayers A and B are
$a_{i2}^{A}=-0.3$ and $a_{i2}^{B}=0$ or $0.2$, respectively.

Asymmetric interface conditions are found to have a significant
effect on the resonance spectrum. As the asymmetry becomes
stronger, the relative intensities of the quasi-antisymmetric
modes increase, while those of the quasi-symmetric modes
decrease, but the critical angle value remains unchanged and
equal to $54.7^{o}$, as in the case of symmetric bilayer film.
This is due to the fact that in the critical configuration a
bilayer film must be fully symmetric.

Thus, in a ferromagnetic bilayer film with 'natural' surfaces and
the interface properties defined by two parameters only, namely
the interface coupling and the interface uni-axial anisotropy,
the critical angle value is always $54.7^{o}$, for both symmetric
and asymmetric interface conditions.

\subsection{The effect of uni-directional anisotropy}

% Fig. 11
\begin{figure}
  \centering
  \includegraphics{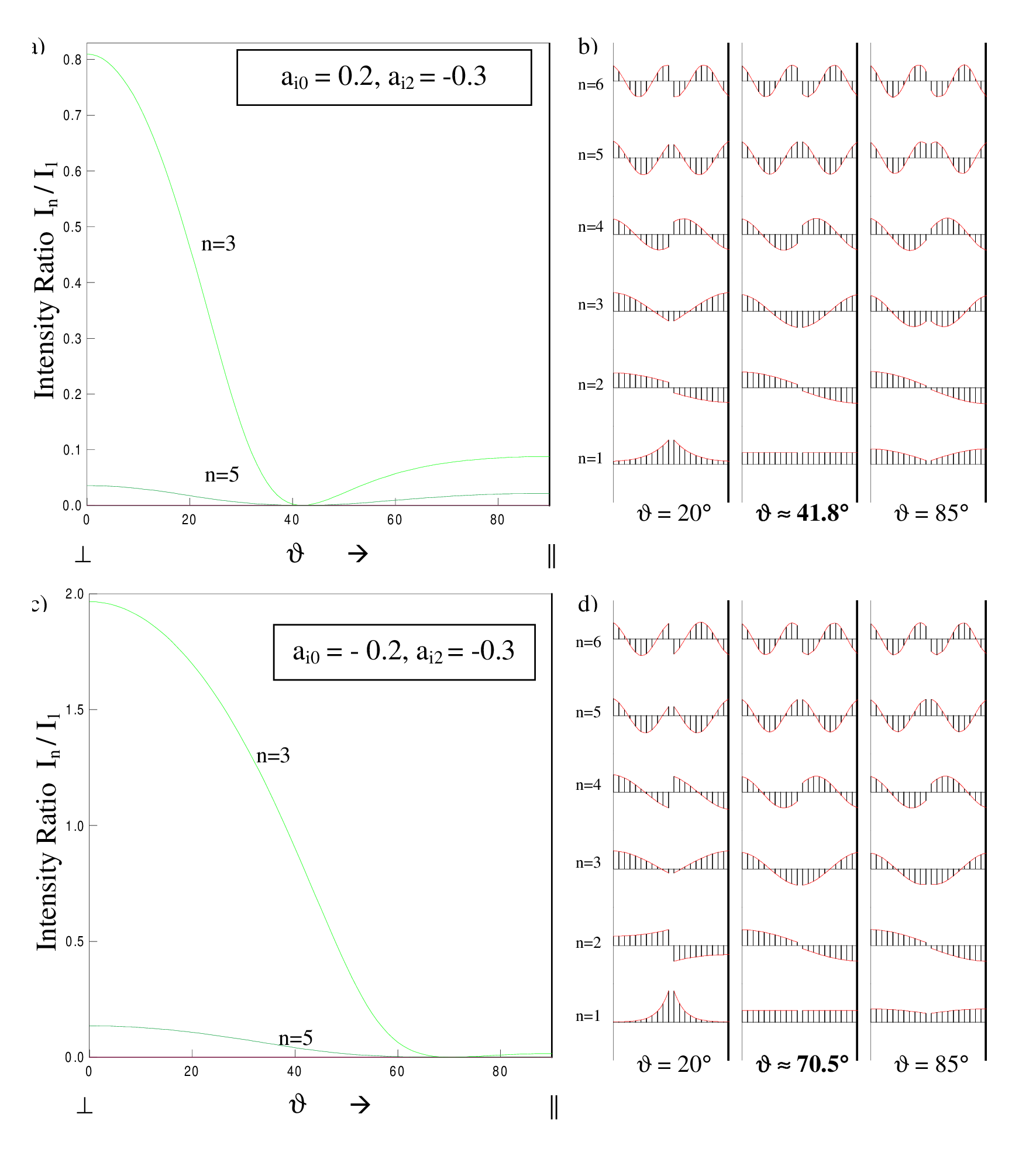}
  %\vspace{20cm}
\caption[]{Resonance line relative intensities {\it versus} the
configuration angle in a bilayer film with symmetric interface
boundary conditions and ferromagnetic coupling, assuming non-zero
uni-directional anisotropy ($J_{int}=0.3$, $a_{i0} \neq 0$). The
other parameter values assumed: $N=11$, $A_{surf}=1$. Graphs (a)
and (c) show intensity ratio $I_{n}/I_{1}$ plotted versus angle
$\theta$ for two different values of $a_{i0}$; the corresponding
profiles of the six lowest modes are shown in (b) and (d) for
three different $\theta$ values.}\label{rys teta 2D}
\end{figure}

\hspace*{\parindent} The effect of the uni-directional anisotropy
on the critical angle value in symmetric bilayer SWR spectrum is
illustrated in Fig. \ref{rys teta 2D}. Each sublayer contains 11
atomic planes, and the coupling between them is of ferromagnetic
nature ($J_{int}=0.3$). The assumed values of interface uni-axial
and uni-directional anisotropies are $a_{i2}=-0.3$ and
$a_{i0}=0.2$; $-0.2$, respectively.

% Fig. 12
\begin{figure}
  \centering
  \includegraphics{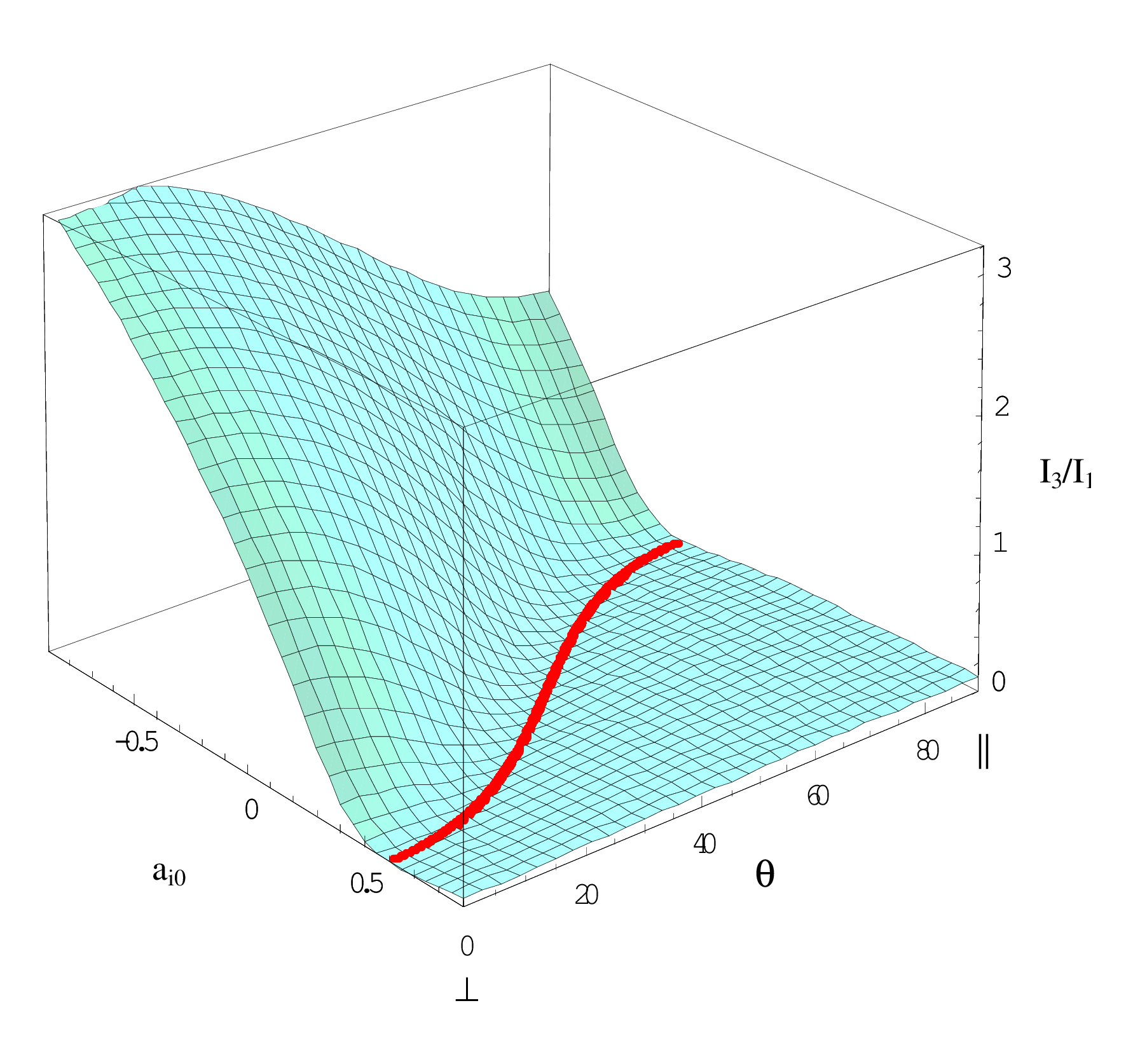}
  %\vspace{17cm}
\caption[]{Intensity ratio $I_{3}/I_{1}$ of the two lowest
symmetric modes (n = 1, 3) in a symmetric bilayer SWR spectrum,
plotted {\it versus} the configuration angle (angle between the
film magnetization and the surface normal), $\theta$, and the
interface uni-directional anisotropy, $a_{i0}$. The ferromagnetic
interface exchange coupling is assumed, with interface uni-axial
anisotropy value $a_{i2}= -0.3$. The bold line represents the
critical angle, $\theta_{crit}$, as a function of the interface
uni-directional anisotropy, $a_{i0}$.}\label{rys teta 3D}
\end{figure}

The critical angle is found to strongly depend on the
uni-directional anisotropy. The exact relation, deduced from
(\ref{param eff int teta}), is as follows:
\begin{equation}\label{teta kryt}
\theta_{crit}=\arccos\sqrt{\frac{1}{3}\left(1-\frac{a_{i0}}{a_{i2}}\right)}.
\end{equation}
The above relation is plotted in Fig. \ref{rys teta 3D} (bold
line); in the same graph, intensity ratio $I_{3}/I_{1}$ is
plotted {\it versus} angle $\theta$ and uni-directional
anisotropy $a_{i0}$. Function (\ref{teta kryt}) follows exactly
the line along which the value of function $I_{3}/I_{1}(\theta
,a_{i0})$ vanishes.

Thus, a critical angle value different from $54.7^{o}$, found
experimentally in a symmetric bilayer SWR spectrum, would provide
evidence for co-existence of uni-axial and uni-directional
anisotropies in the studied sample, the ratio of these two
anisotropy types being deducible from the critical angle value
(according to (\ref{teta kryt})).

\section{Conclusions}\label{rozdz wnioski}

\hspace*{\parindent} The results of our theoretical investigation
of bilayer FMR spectrum, presented in this study, deny the
interpretation commonly used in experimental studies reporting
double-peak FMR spectra in bilayer films. In this interpretation
the low-intensity peak is related to an optic mode, and the
high-intensity line to an acoustic mode. However, our theoretical
study shows this is not always true, providing examples of
double-peak spectra in which {\it both} appearing modes are
acoustic or, what is more, the {\it high-intensity} peak
corresponds to an {\it optic} mode. Thus, the interpretation of
the observed modes cannot be unambiguous {\it a priori}, as the
intensity of a resonance line depends not only on the phase shift
in the sublayer magnetization precession, but also on the
precession amplitude distribution, and especially on its
localization.

In the investigated {\it symmetric} bilayer SWR spectra both the
position and the intensity of the resonance lines are found to be
{\it independent} of the interface coupling integral value. The
reason is that {\it in symmetric bilayer film} the effective
interface parameter depends on the interface coupling integral
for antisymmetric modes only, and these modes {\it do not appear}
in the SWR spectrum. Hence, symmetric bilayer film seems to be
particularly convenient for interface anisotropy studies.
However, it should be remembered that when the interface coupling
is ferromagnetic the odd modes are symmetric, and the even modes
are antisymmetric, while in the case of antiferromagnetic
coupling the situation is reversed: the odd modes (including the
first one, with the lowest energy) are antisymmetric, the even
modes being symmetric.

Moreover, we show that the critical angle effect can occur in a
bilayer film even in the absence of surface anisotropy. In this
case, this effect is totally due to the interface, and if the
interface anisotropy is purely {\it uni-axial}, the critical
angle value is always $54.7^{o}$ and does not depend on the
anisotropy value. Any deviation of the critical angle value from
$54.7^{o}$ involves an additional source of interface anisotropy,
namely the {\it uni-directional} anisotropy.

\begin{center}
ACKNOWLEDGEMENTS
\end{center}

This work was supported by the Polish State Committee for
Scientific Research through projects: KBN-2PO3B 120 23 and
PBZ-KBN-044/PO3-2001.

\addcontentsline{toc}{section}{Appendix: Bilayer Hamiltonian
matrix elements}

\section*{Appendix: Bilayer Hamiltonian matrix elements}%\label{rozdz dod}

\hspace*{\parindent} Below we specify the explicit form of
Hamiltonian matrix elements for cubic crystal surface cuts (001)
and (110). Our formulae were derived for $\vkr =0$, with the
quantization vectors in both sublayers assumed to be identical
({\it i.e.} $\vec{\gamma}_{A} = \vec{\gamma}_{B}$, which implies
$\vec{A}_{A} = \vec{A}_{B}$), and the spin precession to be
circular. On these assumptions matrices (\ref{XA mac}), (\ref{XB
mac}) and (\ref{XAB mac}) become:

\begin{displaymath}
X_{A}=\left[
\begin{array}{ccccc}
R_{A}-a & C_{A} & & & \\
C_{A}^{\ast} & R_{A} & C_{A} & & \\
& \ddots & \ddots & \ddots & \\
& & C_{A}^{\ast} & R_{A}& C_{A} \\
& & & C_{A}^{\ast} & R_{A}-b
\end{array}
\right]; X_{B}=\left[
\begin{array}{ccccc}
R_{B}-c & C_{B} & & & \\
C_{B}^{\ast} & R_{B} & C_{B} & & \\
& \ddots & \ddots & \ddots & \\
& & C_{B}^{\ast} & R_{B}& C_{B} \\
& & & C_{B}^{\ast} & R_{B}-d
\end{array}
\right];
\end{displaymath}
\begin{displaymath}
X_{AB}=\left[
\begin{array}{ccc}
\vdots  & \vdots & \\
   0    & 0 & \ldots \\
C_{int} & 0 & \ldots
\end{array}
\right].
\end{displaymath}

\hspace*{-\parindent} In all the relations detailed below index
$i$ denotes the sublayer label (A or B).

\subsubsection*{sc(001) surface cut}

\begin{eqnarray}\label{el s0}
R_{i} & = & 4S_{i}J_{i} +
  g\mu_{\beta}(\vec{H}_{i}^{eff}\cdot\vec{\gamma})
  + D_{i}\left(S_{i}-\frac{1}{2}\right)
  \left(3\cos^{2}\theta-1\right) , \nonumber \\
C_{i} & = & -2S_{i}J_{i}, \nonumber \\
C_{int} & = & -2\sqrt{S_{A}S_{B}}J_{int} , \nonumber \\
a & = & 2S_{A}J_{A} - g\mu_{B}\vec{K}_{s}^{A}\cdot\vec{\gamma} -
  (D_{s}^{A}-D_{A})(S_{A}-\frac{1}{2})(3\cos^{2}\theta-1) ,
  \nonumber \\
b & = & 2S_{A}J_{A} - 2S_{B}J_{int} -
  g\mu_{B}\vec{K}_{int}^{A}\cdot\vec{\gamma} -
  (D_{int}^{A}-D_{A})(S_{A}-\frac{1}{2})(3\cos^{2}\theta-1),\nonumber\\
c & = & 2S_{B}J_{B} - 2S_{A}J_{int} -
  g\mu_{B}\vec{K}_{int}^{B}\cdot\vec{\gamma} -
  (D_{int}^{B}-D_{B})(S_{B}-\frac{1}{2})(3\cos^{2}\theta-1)
  , \nonumber \\
d & = & 2S_{B}J_{B} - g\mu_{B}\vec{K}_{s}^{B}\cdot\vec{\gamma} -
  (D_{s}^{B}-D_{B})(S_{B}-\frac{1}{2})(3\cos^{2}\theta-1).
\end{eqnarray}

\subsubsection*{bcc(001) surface cut}

\begin{eqnarray}\label{el b0}
R_{i} & = & 16S_{i}J_{i} +
  g\mu_{\beta}(\vec{H}_{i}^{eff}\cdot\vec{\gamma})
  + D_{i}\left(S_{i}-\frac{1}{2}\right)
  \left(3\cos^{2}\theta-1\right) , \nonumber \\
C_{i} & = & -8S_{i}J_{i} , \nonumber \\
C_{int} & = & -8\sqrt{S_{A}S_{B}}J_{int} , \nonumber \\
a & = & 8S_{A}J_{A} - g\mu_{B}\vec{K}_{s}^{A}\cdot\vec{\gamma} -
  (D_{s}^{A}-D_{A})(S_{A}-\frac{1}{2})(3\cos^{2}\theta-1) ,
  \nonumber \\
b & = & 8S_{A}J_{A} - 8S_{B}J_{int} -
  g\mu_{B}\vec{K}_{int}^{A}\cdot\vec{\gamma} -
  (D_{int}^{A}-D_{A})(S_{A}-\frac{1}{2})(3\cos^{2}\theta-1),\nonumber\\
c & = & 8S_{B}J_{B} - 8S_{A}J_{int} -
  g\mu_{B}\vec{K}_{int}^{B}\cdot\vec{\gamma} -
  (D_{int}^{B}-D_{B})(S_{B}-\frac{1}{2})(3\cos^{2}\theta-1)
  , \nonumber \\
d & = & 8S_{B}J_{B} - g\mu_{B}\vec{K}_{s}^{B}\cdot\vec{\gamma} -
  (D_{s}^{B}-D_{B})(S_{B}-\frac{1}{2})(3\cos^{2}\theta-1).
\end{eqnarray}

\subsubsection*{fcc(001) surface cut}

\begin{eqnarray}\label{el f0}
R_{i} & = & 16S_{i}J_{i} +
  g\mu_{\beta}(\vec{H}_{i}^{eff}\cdot\vec{\gamma})
  + D_{i}\left(S_{i}-\frac{1}{2}\right)
  \left(3\cos^{2}\theta-1\right) , \nonumber \\
C_{i} & = & -8S_{i}J_{i} , \nonumber \\
C_{int} & = & -8\sqrt{S_{A}S_{B}}J_{int} , \nonumber \\
a & = & 8S_{A}J_{A} - g\mu_{B}\vec{K}_{s}^{A}\cdot\vec{\gamma} -
  (D_{s}^{A}-D_{A})(S_{A}-\frac{1}{2})(3\cos^{2}\theta-1) ,
  \nonumber \\
b & = & 8S_{A}J_{A} - 8S_{B}J_{int} -
  g\mu_{B}\vec{K}_{int}^{A}\cdot\vec{\gamma} -
  (D_{int}^{A}-D_{A})(S_{A}-\frac{1}{2})(3\cos^{2}\theta-1),\nonumber\\
c & = & 8S_{B}J_{B} - 8S_{A}J_{int} -
  g\mu_{B}\vec{K}_{int}^{B}\cdot\vec{\gamma} -
  (D_{int}^{B}-D_{B})(S_{B}-\frac{1}{2})(3\cos^{2}\theta-1)
  , \nonumber \\
d & = & 8S_{B}J_{B} - g\mu_{B}\vec{K}_{s}^{B}\cdot\vec{\gamma} -
  (D_{s}^{B}-D_{B})(S_{B}-\frac{1}{2})(3\cos^{2}\theta-1).
\end{eqnarray}

\subsubsection*{sc(110) surface cut}

\begin{eqnarray}\label{el s1}
R_{i} & = & 8S_{i}J_{i} +
  g\mu_{\beta}(\vec{H}_{i}^{eff}\cdot\vec{\gamma})
  + D_{i}\left(S_{i}-\frac{1}{2}\right)
  \left(3\cos^{2}\theta-1\right) , \nonumber \\
C_{i} & = & -4S_{i}J_{i} , \nonumber \\
C_{int} & = & -4\sqrt{S_{A}S_{B}}J_{int} , \nonumber \\
a & = & 4S_{A}J_{A} - g\mu_{B}\vec{K}_{s}^{A}\cdot\vec{\gamma} -
  (D_{s}^{A}-D_{A})(S_{A}-\frac{1}{2})(3\cos^{2}\theta-1) ,
  \nonumber \\
b & = & 4S_{A}J_{A} - 4S_{B}J_{int} -
  g\mu_{B}\vec{K}_{int}^{A}\cdot\vec{\gamma} -
  (D_{int}^{A}-D_{A})(S_{A}-\frac{1}{2})(3\cos^{2}\theta-1),\nonumber\\
c & = & 4S_{B}J_{B} - 4S_{A}J_{int} -
  g\mu_{B}\vec{K}_{int}^{B}\cdot\vec{\gamma} -
  (D_{int}^{B}-D_{B})(S_{B}-\frac{1}{2})(3\cos^{2}\theta-1)
  , \nonumber \\
d & = & 4S_{B}J_{B} - g\mu_{B}\vec{K}_{s}^{B}\cdot\vec{\gamma} -
  (D_{s}^{B}-D_{B})(S_{B}-\frac{1}{2})(3\cos^{2}\theta-1).
\end{eqnarray}

\subsubsection*{bcc(110) surface cut}

\begin{eqnarray}\label{el b1}
R_{i} & = & 4S_{i}J_{i} +
  g\mu_{\beta}(\vec{H}_{i}^{eff}\cdot\vec{\gamma})
  + D_{i}\left(S_{i}-\frac{1}{2}\right)
  \left(3\cos^{2}\theta-1\right) , \nonumber \\
C_{i} & = & -4S_{i}J_{i} , \nonumber \\
C_{int} & = & -4\sqrt{S_{A}S_{B}}J_{int} , \nonumber \\
a & = & 4S_{A}J_{A} - g\mu_{B}\vec{K}_{s}^{A}\cdot\vec{\gamma} -
  (D_{s}^{A}-D_{A})(S_{A}-\frac{1}{2})(3\cos^{2}\theta-1) ,
  \nonumber \\
b & = & 4S_{A}J_{A} - 4S_{B}J_{int} -
  g\mu_{B}\vec{K}_{int}^{A}\cdot\vec{\gamma} -
  (D_{int}^{A}-D_{A})(S_{A}-\frac{1}{2})(3\cos^{2}\theta-1),\nonumber\\
c & = & 4S_{B}J_{B} - 4S_{A}J_{int} -
  g\mu_{B}\vec{K}_{int}^{B}\cdot\vec{\gamma} -
  (D_{int}^{B}-D_{B})(S_{B}-\frac{1}{2})(3\cos^{2}\theta-1)
  , \nonumber \\
d & = & 4S_{B}J_{B} - g\mu_{B}\vec{K}_{s}^{B}\cdot\vec{\gamma} -
  (D_{s}^{B}-D_{B})(S_{B}-\frac{1}{2})(3\cos^{2}\theta-1).
\end{eqnarray}

\end{document}